\documentclass[prb,reprint,amsmath,amssymb,floatfix,superscriptaddress,nofootinbib]{revtex4-2}
\usepackage{graphicx}
\usepackage{dcolumn}
\usepackage{bm}
\usepackage{xcolor}
\usepackage{braket}
\usepackage[normalem]{ulem}
\usepackage{cancel}

\newcommand{\stkout}[1]{\ifmmode\text{\sout{\ensuremath{#1}}}\else\sout{#1}\fi}

\begin{document}
\title{Exploring Spin AGP Ansatze for Strongly Correlated Spin Systems}

\author{Zhiyuan Liu}
\affiliation{Department of Physics and Astronomy, Rice University, Houston, TX 77005-1892}

\author{Fei Gao}
\affiliation{Department of Physics and Astronomy, Rice University, Houston, TX 77005-1892}

\author{Guo P. Chen}
\affiliation{Department of Chemistry, Rice University, Houston, TX 77005-1892}

\author{Thomas M. Henderson}
\affiliation{Department of Physics and Astronomy, Rice University, Houston, TX 77005-1892}
\affiliation{Department of Chemistry, Rice University, Houston, TX 77005-1892}

\author{Jorge Dukelsky}
\affiliation{Instituto de Estructura de la Materia, IEM-CSIC, Serrano 123, 28006 Madrid, Spain}

\author{Gustavo E. Scuseria}
\affiliation{Department of Physics and Astronomy, Rice University, Houston, TX 77005-1892}
\affiliation{Department of Chemistry, Rice University, Houston, TX 77005-1892}

\date{\today}

\begin{abstract}
The antisymmetrized geminal power (AGP), a wave function equivalent to number-projected Hartree--Fock--Bogoliubov (HFB), and number-projected Bardeen--Cooper--Schrieffer (BCS) when working in the paired (natural orbitals) basis, has proven to be an excellent reference for strong pairing interactions. Several correlation methods have also been applied on top of AGP. In this work, we show how AGP can also be applied to spin systems by simply basing its formulation on a spin $su(2)$ algebra. We here implement spin AGP and spin AGP-based correlation techniques and benchmark them on the XXZ and $\mathrm{J_1-J_2}$  Heisenberg models, both in 1 and 2 dimensions. Our results indicate that spin AGP is a promising starting point for modeling spin systems.
\end{abstract}

\maketitle

\section{Introduction}
Model spin Hamiltonians provide valuable insight into magnetic materials, high-temperature superconductors, and biochemical processes such as nitrogen fixation \cite{mikeska2004one,science.abm2295,NOODLEMAN1995199}. They are also important for the study of quantum sensors, cold atoms in optical lattices, and fault-tolerant quantum computers \cite{davis2021probing,PhysRevLett.93.250405,science.aay0668}. These model Hamiltonians capture diverse physical phenomena without the details of a fully \textit{ab initio} description. Nevertheless, with a few exceptions \cite{KITAEV20062}, lattice models of spin systems beyond one dimension are not exactly solvable, and we have to resort to approximate numerical methods.

Here we focus on the ground states of spin lattice models, whose computation is already challenging due to various quantum phases that arise from different interaction strengths \cite{mikeska2004one,science.1201080,PhysRevLett.99.127004,PhysRevB.91.081104,PhysRevB.92.041105,LIU20221034}.   Particularly, analogous to Hartree--Fock in electronic structure theory, spin wave functions based on a single spin configuration are inadequate in the strongly correlated regime \cite{bishop1992coupled,bishop1996coupled}.  However, our recent work suggests that methods in electronic structure theory can be useful for studying spins if they are mapped to fermions without constraints \cite{doi:10.1063/5.0125124}.

The antisymmetrized geminal power (AGP) wave function \cite{Coleman1965,ring2004nuclear} has been shown to be a good starting point for certain strongly correlated problems. When correlated with configuration interaction (CI) or coupled cluster (CC) theory \cite{henderson2019geminal,henderson2020correlating}, AGP yields quite accurate results for the reduced Bardeen--Cooper--Schrieffer (BCS) Hamiltonian, which models the kinds of strong correlations seen in conventional superconductors \cite{PhysRev.108.1175,RevModPhys.34.694}.

Though AGP was originally developed for paired fermionic systems, the pairing algebra generators satisfy the same $su(2)$ algebra as spin operators.  Inspired by Anderson's resonating valence bond theory which was applied to study both the Heisenberg model and Hubbard model \cite{anderson1973resonating,anderson1987resonating}, we propose to treat spin systems via AGP.  Our results suggest that spin AGP (sAGP) and correlated methods based on it are computationally affordable techniques that can accurately describe the ground states of strongly correlated spin systems.

\section{Theory}
\subsection{Antisymmetrized Geminal Power}

The central concept of AGP \cite{Coleman1965,ring2004nuclear} is the geminal, a two-electron wave function created by a geminal creation operator
\begin{equation}
  \Gamma^\dagger
  =\frac{1}{2} \sum_{1 \leq p,q \leq 2M} \eta_{pq} \, c_p^\dagger \, c_q^\dagger,
\label{eq:geminal_operator_fermion}
\end{equation}
where $\eta$ is antisymmetric, $c_p^\dagger$ is the fermion creation operator for spinorbital $p$, and indices $p$, $q$ run over all $2M$ spinorbitals. An AGP state with $2N$ electrons is created by occupying the same geminal $N$ times:
\begin{equation}
|\mathrm{AGP}\rangle = \frac{1}{N!} \, \left(\Gamma^\dagger\right)^N |-\rangle,
\label{AGP_P}
\end{equation}
where $|-\rangle$ is the physical vacuum.

In practice, it is more convenient to work in the natural orbital basis of the geminal, where
$\eta$ is quasidiagonal \cite{hua1944theory},
\begin{equation}
  \eta = \bigoplus_{p=1}^M
  \begin{pmatrix}
    0 & \eta_p\\
    -\eta_p &0
  \end{pmatrix},
\end{equation}
displaying a pairing scheme of the spin-orbitals \cite{henderson2020correlating}. In this basis, we can write
\begin{equation}
\Gamma^\dagger=\sum_{p=1}^{M} \eta_{p} \, P_p^\dagger,
\label{eq:geminal_operator}
\end{equation}
in which we have defined
\begin{equation}
  P_p^\dagger=c_p^\dagger \, c_{\bar{p}}^\dagger
\label{P_opr}
\end{equation}
and have reindexed the fermion creation operators by $p$ and its paired companion $\bar{p}$, where $p$ enumerates all $M$ pairs. The AGP then assumes the form of an elementary symmetric polynomial:
\begin{equation}
|\mathrm{AGP}\rangle = \sum_{1 \leq p_1 < \ldots p_N \leq M} \eta_{p_1} \ldots \eta_{p_N} \, P_{p_1}^\dagger \ldots P_{p_N}^\dagger |-\rangle.
\label{AGP_P_expand}
\end{equation}
Because AGP is equivalent to number-projected Hartree--Fock--Bogoliubov (HFB) \cite{scuseria2011projected} or number-projected BCS in the natural orbital basis, it can be optimized with mean-field cost of $\mathcal{O}(M^3)$ \cite{khamoshi2019efficient, dutta2021construction,10.1063/5.0156124}, and its variationally optimized result is guaranteed to be at least as good as Hartree--Fock, which is just a special case of AGP in which only $N$ of the $\eta$ values are non-zero.

In this work, we will not worry about the norm of the AGP wave function, which can be normalized by multiplying all the $\eta$ values by the same constant.

\subsection{AGP for Spin Systems}
The pair creation operator $P_p^\dagger$ and its adjoint $P_p $, together with the number operator
\begin{equation}
  N_p= c_p^\dagger \, c_p + c_{\bar{p}}^\dagger \, c_{\bar{p}},
\label{N opr}
\end{equation}
close the the $su(2)$ commutation algebra:
\begin{subequations}
\label{SU2}
\begin{align}
[P_p,P_q^\dagger] &=\delta_{pq} \, (1-N_p),
\\
[N_p,P_q^\dagger] &= 2\, \delta_{pq} \, P_p^\dagger.
\end{align}
\end{subequations}

\begin{figure*}
\includegraphics[width=\columnwidth]{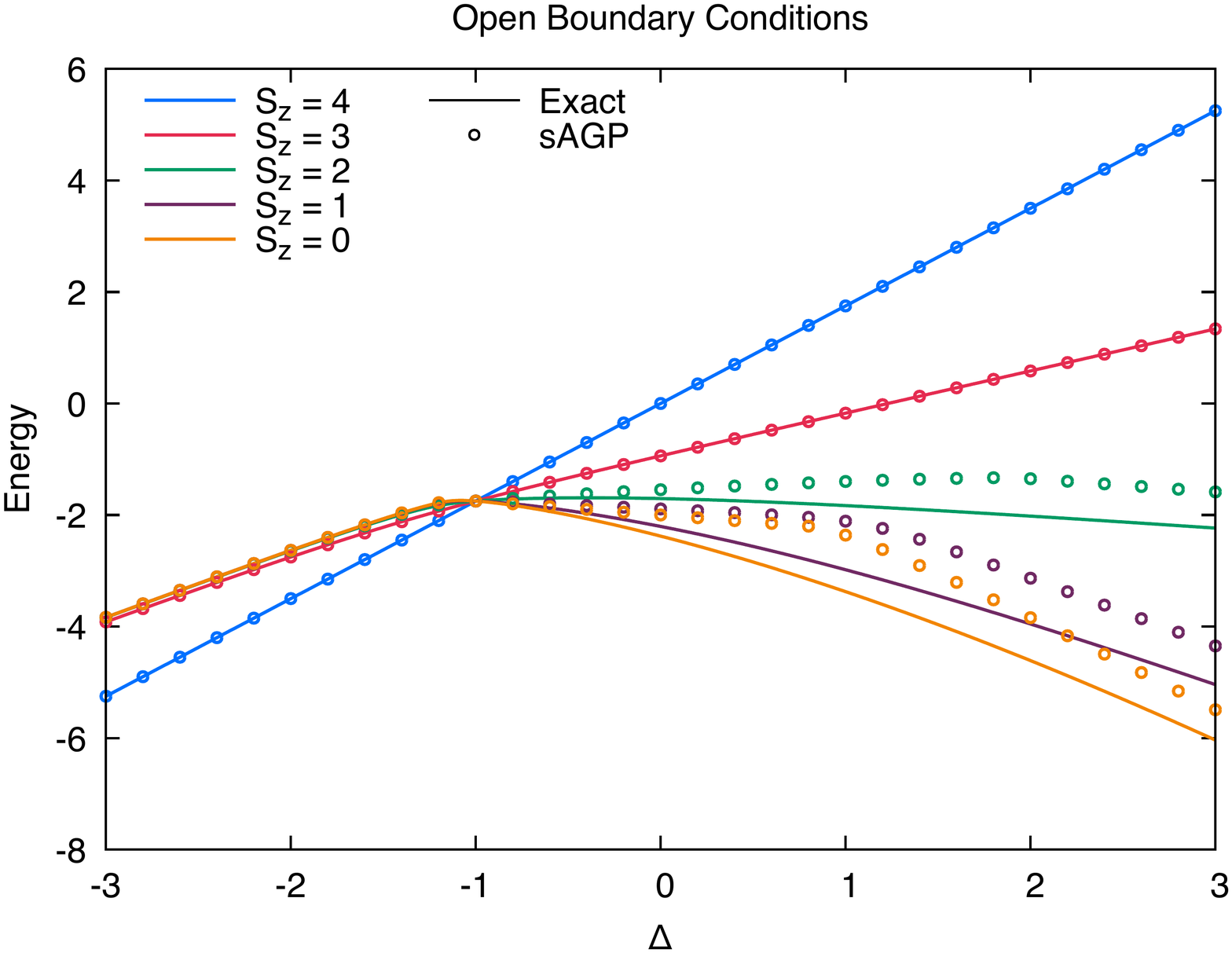}
\hfill
\includegraphics[width=\columnwidth]{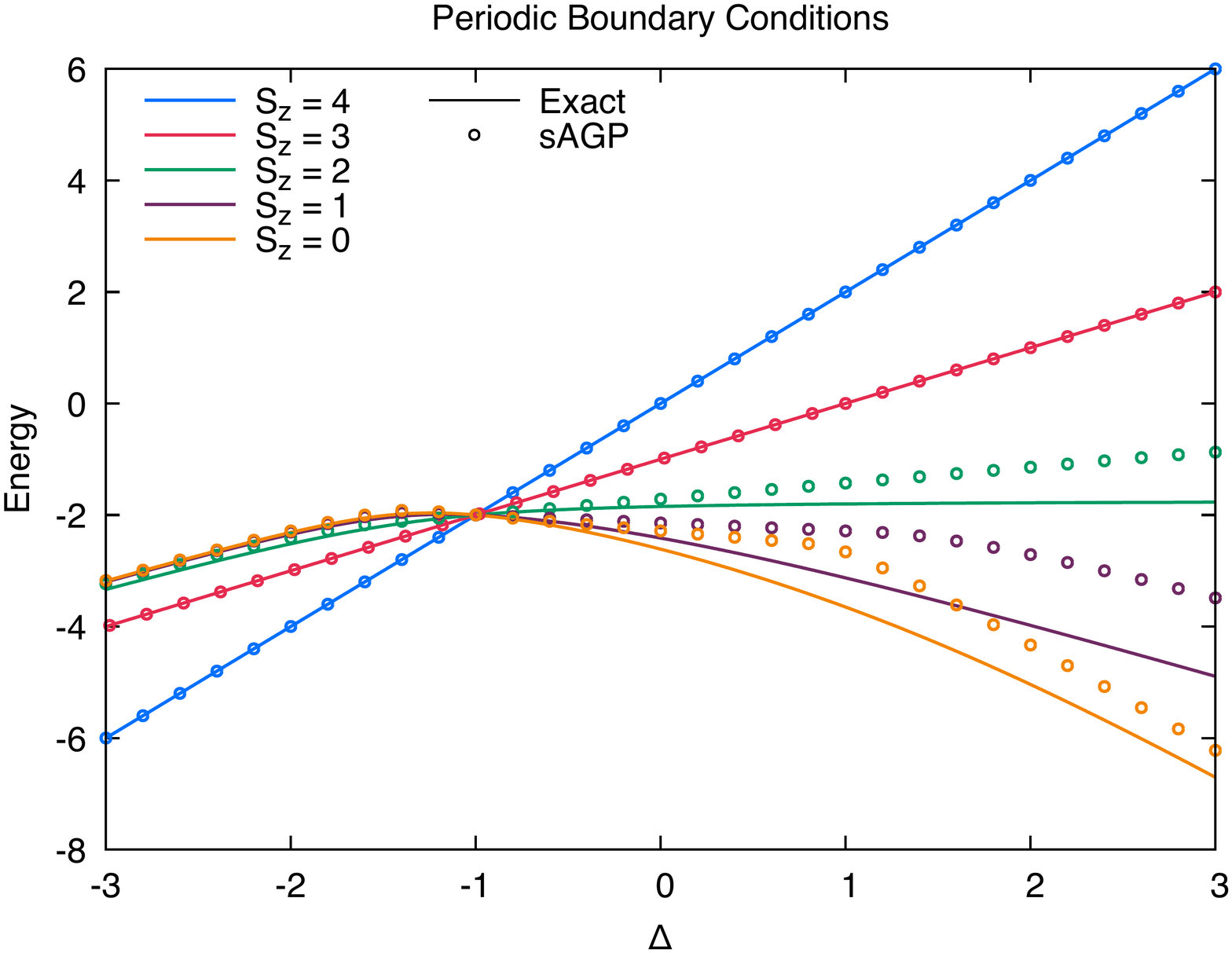}
\caption{Energies in the 8-site 1D XXZ Hamiltonian for different $S^z$ sectors and open boundary conditions (left panel) or periodic boundary conditions (right panel).  We compare the exact results (lines) against the mean-field optimized sAGP (points).  Different colors correspond to different $S^z$ sectors.  Spin AGP is very accurate for $\Delta < -1$ and exact for all $S^z$ sectors at $\Delta = -1$. We note that sAGP is always exact for $S^z=3$ and $S^z=4$ as explained in Sec.~\ref{XXZdiffSz}.  The curves for $S^z=0$ and $S^z=1$ are hard to distinguish for $\Delta<0$ in this figure but they are not identical and $S^z=0$ has a higher energy.
\label{Fig:8_sites}}
\end{figure*}

Following Anderson \cite{Anderson1958}, we can relate the AGP commutation algebra to the spin-$\frac{1}{2}$ $su(2)$ 
\begin{subequations}
\begin{align}
[S_p^+,S_q^-] &= 2 \, \delta_{pq} \, S_p^z,
\\
[S_p^z,S_q^+] &= \delta_{pq} \, S_p^+.
\end{align}
\label{SU2_S}
\end{subequations}
Comparing with Eqns.~\eqref{SU2}, we see that by the bijective mapping
\begin{subequations}
\begin{align}
S_p^+ &\leftrightarrow P_p^\dagger,
\\
S_p^- &\leftrightarrow P_p,
\\
S_p^z &\leftrightarrow \frac{N_p - 1}{2},
\end{align}
\label{s_p_map}
\end{subequations}
we can simply transcribe any expressions for standard AGP matrix elements in the zero seniority \cite{bytautas2011seniority} fermion space, where all electrons are paired, to those for spin AGP (sAGP for short), and can readily generalize any of the techniques we have introduced for the correlation of AGP to sAGP \cite{henderson2019geminal,henderson2020correlating,dutta2020geminal,dutta2021construction,khamoshi2021exploring}.
In the standard pairing AGP case,
we have
\begin{equation}
  P_p |-\rangle=0,
\end{equation}
where $|-\rangle$ denotes the physical vacuum.
The corresponding spin ``vacuum" state $\ket{\Downarrow}$ is the product state of $\downarrow$-spins on all sites, and statisfies
\begin{equation}
  S_p^- \ket{\Downarrow} =0.
\end{equation}

The sAGP wave function is thus
\begin{subequations}
\begin{align}
  |\mathrm{sAGP}\rangle &= \frac{1}{N!} \, \left(\Gamma^\dagger\right)^N \ket{\Downarrow},
\\
\Gamma^\dagger &= \sum_p \eta_p \, S_p^+,
\end{align}
\label{sAGP}
\end{subequations}
where we have a total of $N$ $\uparrow$-spins and $(M-N)$ $\downarrow$-spins, so
\begin{equation}
  \langle \mathrm{sAGP}|S^z|\mathrm{sAGP}\rangle = N - \frac{M}{2}.
\end{equation}
At half filling ($N = M/2$) the sAGP wave function is magnetically neutral.

Incidentally, the inverse mapping of Eqns.~\eqref{s_p_map} has been used to implement quantum computing algorithms for the standard pairing AGP state \cite{khamoshi2020correlating, khamoshi2023quantum, doi:10.1021/acs.jpca.3c00525}.

Let $\ket{p_1p_2\cdots p_N}$ be a spin product state (SPS) where the spins are up on sites $p_1,p_2,\cdots p_N$ and down on the others. Eqns.~\eqref{sAGP} implies
that sAGP is a linear combination of all SPSs in the Hilbert space of the problem,
with coefficients
\begin{equation}
  \langle p_1p_2\cdots p_N|\mathrm{sAGP}\rangle = \eta_{p_1}\eta_{p_2}\cdots\eta_{p_N}.
\end{equation}
This means that sAGP is a particularly simple matrix product state, whose matrix elements are straightforward and inexpensive to compute \cite{khamoshi2019efficient, dutta2021construction,10.1063/5.0156124}.

We have noted that standard AGP is equivalent to number-projected BCS, which suggests that there should be an equivalent projected mean-field understanding of sAGP.  This is indeed the case: sAGP is simply the $S^z$-projected spin BCS state, where spin BCS (sBCS) is defined as
\begin{equation}
\left\vert \mathrm{sBCS}\right\rangle =\prod\limits_{p=1}^{M}\left( 1+\eta _{p}S_{p}^{+}\right)
\ket{\Downarrow},
\end{equation}
in analogy with the standard BCS given in terms of pairing operators $P_p^\dagger$ and the physical vacuum.  When the spin problem is mapped to fermions, spin BCS corresponds to generalized Hartree--Fock (GHF) in which the spin-orbitals break not only $S^2$ but also $S^z$ symmetry \cite{changlani2018macroscopically,pal2021colorful,chertkov2021motif}. 

In this work, in which we specialize to spin Hamiltonians, the GHF wave function has seniority symmetry dictated by the spins, and one could think of sAGP as an $S^z$-projected general spin product state.

\section{Applications}
We benchmark sAGP on two families of spin lattice systems, the XXZ and $\mathrm{J_1-J_2}$ Heisenberg models \cite{mikeska2004one}.  The former captures anisotropic interactions, while the latter includes interactions beyond nearest neighbors.

We focus predominantly on the nearest-neighbor XXZ model.  We start with the one-dimensional (1D) case as a prototypical example that illustrates the most important features of sAGP and is exactly solvable via the Bethe ansatz \cite{yang1966one}.  We then discuss various two-dimensional (2D) XXZ lattices as well as the $\mathrm{J_1-J_2}$ square lattice, which are not integrable in general.

We first explore sAGP on its own for these systems. While sAGP itself is of modest accuracy in general, we want to understand its properties to provide context for the correlated sAGP results, which we then compare with conventional correlation methods to show that sAGP is a better starting point for strongly correlated spin systems.

\subsection{Spin AGP for the One-Dimensional XXZ Model}

The XXZ Hamiltonian is
\begin{subequations}
\label{XXZ} 
\begin{align}
  H_{\mathrm{XXZ}}
  &= \mathrm{J} \, \sum_{\langle pq \rangle} \left(S_p^x \, S_q^x + S_p^y \, S_q^y + \Delta \, S_p^z \, S_q^z\right)
\\
  &= \mathrm{J} \, \sum_{\langle pq \rangle} \left[\frac{1}{2} \, \left(S_p^+ \, S_q^- + S_p^- \, S_q^+\right) + \Delta \, S_p^z \, S_q^z\right], 
\end{align}
\end{subequations}
where $p$ and $q$ index lattice sites and the notation $\langle pq \rangle$ restricts
the summation over $p$ and $q$ to nearest neighbors.  Generally speaking, we take $\mathrm{J} = 1$ in this article unless otherwise specified,
so that the system is antiferromagnetic when $\Delta > 1$.

In the 1D case, sites $p$ and $q$ are nearest neighbors if $|p-q| = 1$.  With $\mathrm{J} > 0$, it exhibits a N\'eel antiferromagnetic phase for $\Delta \gtrsim 1$, and a ferromagnetic phase for $\Delta \lesssim -1$.  In the interval region $|\Delta| \lesssim 1$, the system is in the XY phase characterized by gapless excitations and long range correlations \cite{mikeska2004one}. While the ferromagnetic and antiferromagnetic phases are fairly simple to describe, the XY phase is much more complicated, and methods based on a single spin configuration struggle (see below and also Ref. \cite{bishop1992coupled,bishop1996coupled}). Spin AGP, however, is exact at $\Delta = -1$, which gives us hope that it will be able to accurately describe this challenging phase as $\Delta$ progresses from $-1$ to $+1$.

\subsubsection{Energies for Different $S^z$ Sectors}
\label{XXZdiffSz}
Let us start with an overview of 
the exact and sAGP ground state energies for different $S^z$ quantum numbers and different values of $\Delta$, as shown in Fig.~\ref{Fig:8_sites}.  For $\Delta < -1$, the exact ground state occurs when all the spins are aligned, i.e., at $S^z = \pm M/2$.  For $\Delta > -1$, the exact ground state is instead $S^z = 0$.  At $\Delta = -1$, the different $S^z$ sectors are exactly degenerate. Spin AGP is exact at $\Delta = -1$ for all $S^z$ sectors and is highly accurate for $\Delta < -1$.  For $\Delta > -1$, sAGP is exact for $S^z = \pm\frac{M}{2}$ and $S^z = \pm(\frac{M}{2}-1)$, but shows larger error as we approach half-filling ($S^z = 0$). As a matter of fact, sAGP is always exact at the $S^z = \pm\frac{M}{2}$ and $S^z = \pm(\frac{M}{2}-1)$ sectors as it has sufficient variational flexibility. $S^z = \pm\frac{M}{2}$ corresponds to the state where all the spins are aligned up or down and sAGP naturally capture it by letting $N=0~\mathrm{or}~M$ respectively. $S^z = \pm(\frac{M}{2}-1)$ means the system has only one $\uparrow$-spin (or $\downarrow$-spin), the exact ground state takes the form:
\begin{equation}
  |\Psi\rangle = \sum_{p}c_p S_{p}^{+}\left\vert \Downarrow  \right\rangle,
\end{equation}
which is just the sAGP state with $N=1$.

\subsubsection{Bimodal Extreme sAGP}

\begin{figure}
\includegraphics[width=\columnwidth]{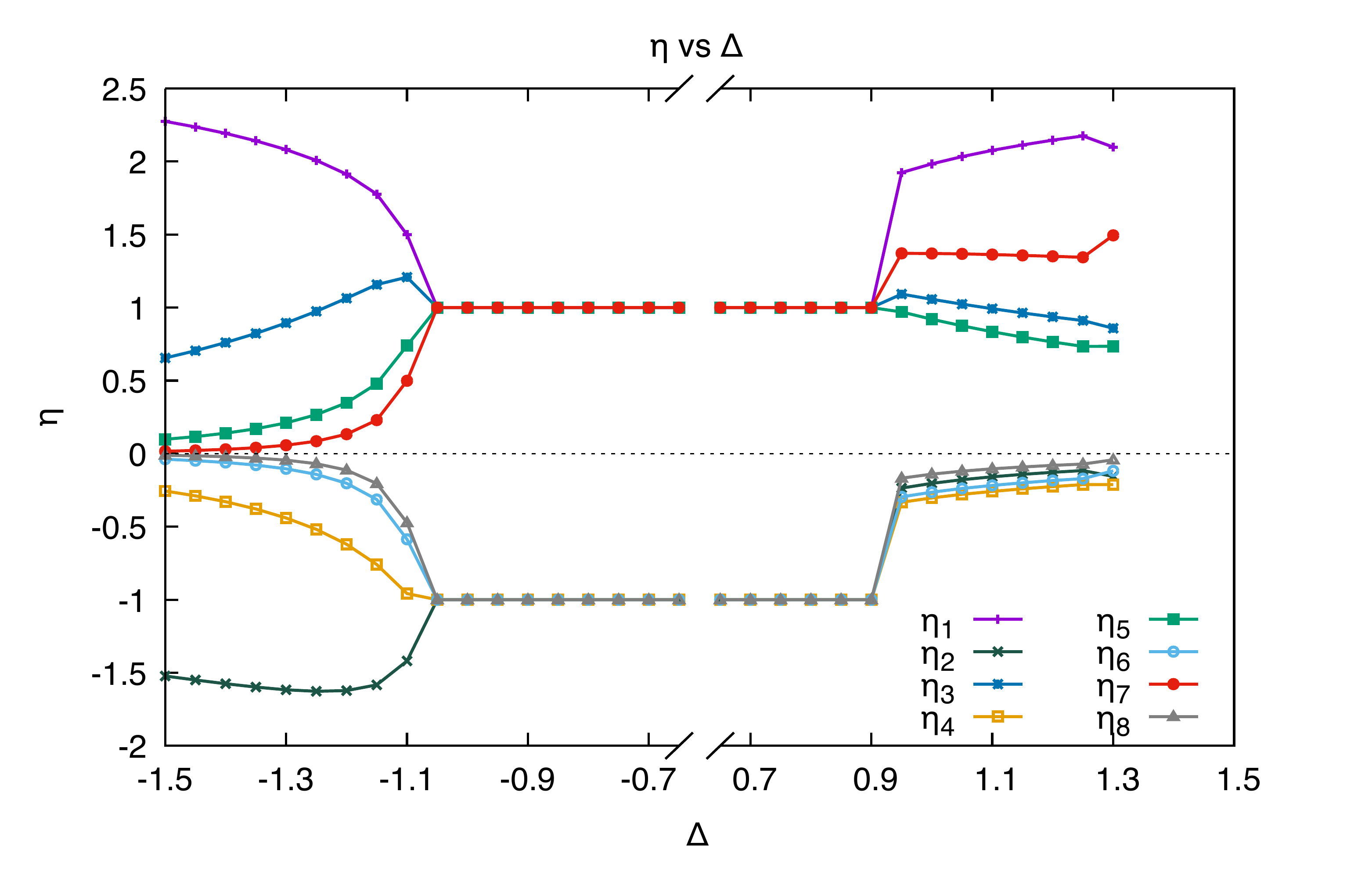}
\caption{ The sAGP geminal coefficient $\eta$ as a function of $\Delta$ for the 8-site XXZ Heisenberg model with open boundary conditions and $ S^z = 0$.  For $-1 \leq \Delta \lesssim 1$ the $\eta$ values are independent of $\Delta$, and $-0.6 < \Delta < 0.6$ has been omitted from the plot.  We order the sites from left to right as $\eta_1$ to $\eta_8$.
\label{Fig:Eta}}
\end{figure}

We now turn to the nature of the sAGP ground state.  We find that $\eta$ values on adjacent sites have opposite signs, for all values of $\Delta$.  When $\Delta$ is large and negative, the $\eta$ values on the left (or right) half of the lattice are large in magnitude, and on the other half are small.  For a site $p$, larger $|\eta_p|$ correspond to larger $\langle S_p^z \rangle$; thus, the fact that the large $|\eta|$ values localize on the left (or right) side of the lattice means that the $\uparrow$ spins localize on this side, i.e, we have a 2-block ferromagnetic solution. Due to the breaking of inversion lattice symmetry, $\uparrow$ spins can either localize on the left half or right, corresponding to two degenerate states. On the other hand, when $\Delta$ is large and positive, alternating sites exhibit a pattern of large and small $|\eta|$, corresponding to a N\'eel arrangement of spins.  These observations are exemplified by the 8-site XXZ chain with open boundary conditions and $S^z = 0$, whose $\eta$ values are depicted in Fig.~\ref{Fig:Eta}.

The more interesting region is of course when $-1 \lesssim \Delta \lesssim 1$, particularly at $\Delta = -1$ where sAGP is exact.  In this region, the sAGP wave function is what we refer to as \textit{bimodal extreme}, for which we find $\eta = (1,-1,1,-1\ldots)$, as can be seen from Fig.~\ref{Fig:Eta}.  An sAGP is extreme when all $\eta$ values are the same in magnitude, which corresponds to each site having equal  $\braket{S^z}$.  We refer to the sAGP as bimodal when the $\eta$ take on two values, in this case, $\pm 1$.  This bimodal extreme sAGP is the exact ground state wave function for $\Delta = -1$ and is the lowest energy sAGP state throughout this XY phase. Note that extreme AGP also has a place in the reduced BCS Hamiltonian $H=\sum_p \epsilon_p N_p -G\sum_{pq} P_p^\dagger P_q$, where, as the interaction strength $G$ goes to infinity, the values of all $\eta$ approach the same \cite{henderson2019geminal}, exhibiting a unimodal extreme AGP that carries off-diagonal long-range order, i.e., superconductivity without number-symmetry breaking \cite{RevModPhys.34.694}.

\
\begin{figure}
  \includegraphics[width=\columnwidth]{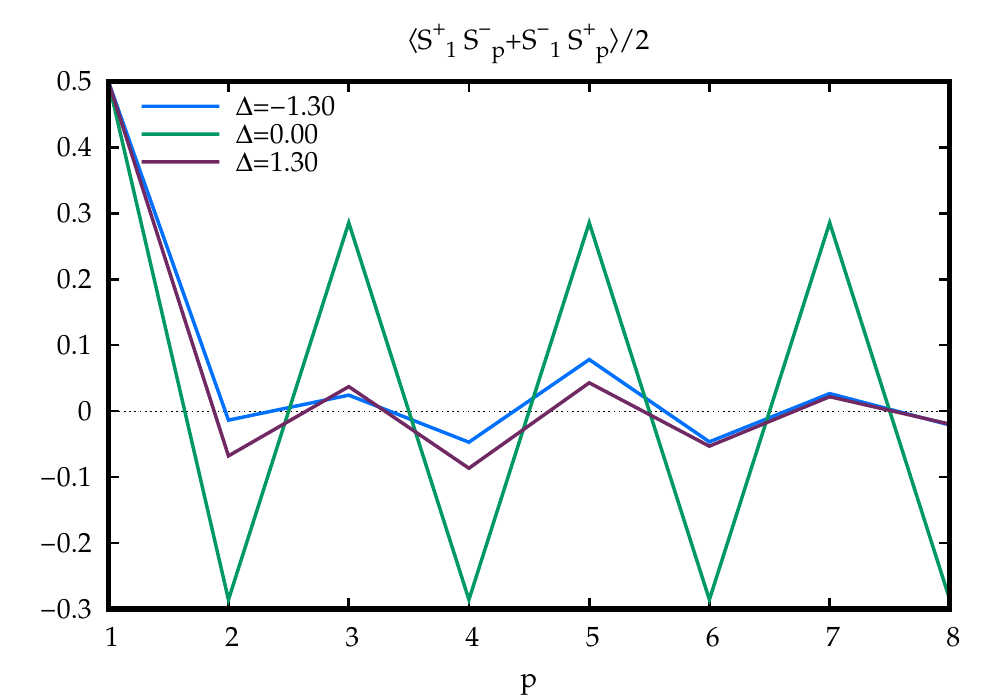}
  \caption{Correlation function $\left < S_1^+S_p^- + S_1^-S_p ^+\right >$/2 for the 8-site XXZ Heisenberg model with open boundary conditions and $ S^z = 0$ for $\Delta = -1.30, 0.00$ and $1.30$, corresponding to the 3 phases of the XXZ model. We see that $\left < S_1^+S_p^- + S_1^-S_p ^+\right >$/2 have alternating signs for even and odd $p$, which is a result of the alternating signs of $\eta_p$.
  \label{Fig:Corr}}
  \end{figure}

We should emphasize again that we do not artificially choose $\eta$ to have a bimodal extreme pattern.  Instead, we variationally optimize the $\eta$ values, and observe that across a wide range of $\Delta$ values, for many different lengths of the XXZ chain and for many different $S^z$ eigenvalues, and for both periodic boundary conditions (PBC) and open boundary conditions (OBC), the variational optimization selects these $\eta$ values.  We also note that bimodal extreme sAGP is always a stationary point of the energy, and the points at which the values of $\eta$ begin to change from extreme occur when it is no longer the lowest energy solution.

Finally, we should say a few words about the physical meaning of the $\eta$ values.  First, we note that the sign of $\eta_p \eta_q$ determines the sign of $\braket{S_p^+S_q^-+S_p^-S_q^+}$ \cite{khamoshi2019efficient}. This can also be seen from Fig.~\ref{Fig:Corr}. If two sites have oppositely-signed $\eta$ values, those sites tend to be antiferromagnetically coupled.  The alternating signs of the $\eta$ values in the bimodal extreme AGP therefore reflect the Marshall sign rule \cite{marshall1955antiferromagnetism}.  The absolute value of $\eta$ on a site, as we can see from Eqn.~\eqref{sAGP}, determines the chance that the spin on that site is flipped to spin-up.  Sites with very large or very small relative $\eta$ values are sites which are not strongly entangled with the other sites.  Sites for which the absolute values of the $\eta$ are similar are more strongly entangled.  The bimodal extreme AGP is actually the maximally entangled state, and in this case each site has $\braket{S^z}=0$.

\subsubsection{Approaching the Thermodynamic Limit}
\begin{figure}
  \includegraphics[width=\columnwidth]{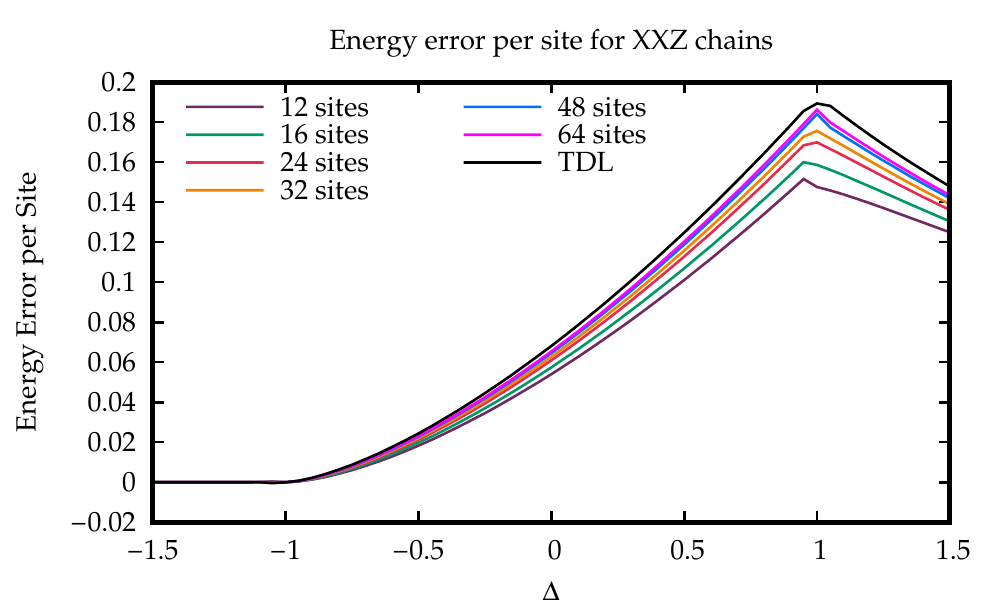}
  \caption{Energy error per site for 1D XXZ chains with different lengths with open boundary conditions. The thermodynamic limit result is obtained by fitting the sAGP energy result by the inverse of the lattice length. We notice that the per-site energy error grows as system size grows for $\Delta>-1$. It is also noticeable for $\Delta<-1$, the per-site energy error reduces as system size grows.
  \label{Fig:TDL}}
  \end{figure}

Fig.~\ref{Fig:TDL} shows the energy error per site for the open boundary XXZ chain with different lengths in the $S^z = 0$ sector.
The energy per site in the thermodynamic limit (TDL), $e_0$, is extrapolated by fitting
\begin{equation}	
	\frac{E(M)}{M}= e_0+e_1\frac{1}{M}+e_2\frac{1}{M^2}+\cdots,
\end{equation}
where we truncate the expansion at second order $e_2 \frac{1}{M^2}$.
We use the same extrapolation scheme for both the sAGP and the exact energies, and display their differences in the TDL in Fig.~\ref{Fig:TDL}.
We observe that for all lattice lengths, sAGP reaches its maximum error around $\Delta = 1$, and the value of $\Delta$ at which the error is the largest grows
with the system size. The maximum sAGP error per site in the TDL is around 0.18. We can also see that sAGP is quite accurate in the ferromagnetic regime ($\Delta<-1$) for all system sizes;
especially, as the system size grows, the per-site error reduces.

\subsubsection{The Ferromagnetic $XXZ$ Model}

So far we have focused on the antiferromagnetic XXZ model, where $\mathrm{J}=1$.
We now briefly discuss the ferromagnetic XXZ model, in which $\mathrm{J} = -1$.
Because of the Hamiltonian's overall sign change, the bimodal extreme sAGP, which is the ground state for the antiferromagnetic XXZ model at $\Delta = -1$,
becomes the highest energy eigenstate at this point for the ferromagnetic XXZ model.
At the Heisenberg point $\Delta=1$, 
an extreme unimodal sAGP where all the $\eta$ values are the same
becomes the ground state for the ferromagnetic XXZ model for all $S^z$ sectors with an energy of $E=-\frac{1}{4}M$ or $E=-\frac{1}{4}(M-1)$
for periodic boundary conditions and open boundary conditions, respectively.

\subsection{Spin $su(2)$ Algebras and Multimodal Extreme sAGPs}
\label{sec:multimodal}

The bimodal extreme sAGP for the 1D antiferromagnetic XXZ model and the unimodal extreme sAGP for the ferromagnetic XXZ model as mentioned above are just two special cases of multimodal extreme sAGPs,
all of which can be formed from collective spin operators which realize a collective $su(2)$ algebra:
\begin{equation}
	K^{\pm }_k=\sum_{p}e^{\pm ikp}S_{p}^{+}\text{, \ }K^{z}=\sum_{p}S_{p}^{z}=S^{z}
	\text{, }  \label{Koperators}
\end{equation}
where $k$ is the lattice momentum. In 1D $k=\frac{2 \pi n}{M}$ with $n$ being an integer restricted to $-\frac{M}{2} < n \leq \frac{M}{2}$. These three operators fulfill the
$su(2)$ commutation algebra
\begin{subequations}
\begin{align}
 \left[ K^{+}_k,K^{-}_k\right]
 &= 2K^{z},\\
 \left[K^{z},K^{\pm}_k\right]
 &=\pm K^{\pm}_k
\end{align}
\end{subequations}
Note that for momentum $k=0$ the $K$-$su(2)$ algebra reduces to the spin $su(2)$ algebra. 

This $K$-$su(2)$ algebra has been recently introduced in the context of quantum many-body scars in spin lattice systems \cite{PhysRevResearch.2.043305,PhysRevResearch.4.013155}. However, our goal here is to use it to construct a reference ansatz to study many-body correlations in spin lattice ground states. 

The (unnormalized) $K$-spin extreme sAGP state is a $K_k$-spin-$\frac{M}{2}$ multiplet,
\begin{equation}
	\left\vert N_k \right\rangle =\left( K^{+}_k\right) ^{N} \left\vert \Downarrow
	\right\rangle , 
	\label{N}
\end{equation}
with $K^{z}=N-\frac{M}{2}$ and $K_k^{2}=\frac{M}{2}\left( \frac{M}{2}+1\right)$.
Note that each site has the same $\braket{S^z}$ in this wave function.
The special cases $k = \frac{2\pi}{m}$ for integer $m$ constitute the $m$-modal extreme AGP states. In these cases we have
\begin{equation}
	\left\vert N_k \right\rangle =\left(\sum_{p}e^{i\frac{2\pi p}{m}}S_{p}^{+}\right) ^{N} \left\vert \Downarrow
	\right\rangle. 
	\label{N}
\end{equation}
One can see that the $\eta$ values are the $m^{th}$ roots of unity.
For $m = 1, 2, 3$,  the $m$-modal extreme AGP states are specifically called unimodal, bimodal, trimodal extreme AGP, respectively.
These $m$-modal extreme AGP states are a special class of AGP states, which, as we see here, are the  ${K_{2\pi/m}}$-spin eigenstates.

We can now ask under what conditions the $m$-modal extreme AGP states $\left\vert N_k \right\rangle$
are eigenstates of the XXZ Hamiltonian. As demonstrated in \cite{pal2021colorful},
it depends on the geometry of the lattice.
For the 1D XXZ Hamiltonian with PBC, the condition is  
\begin{equation}
  \Delta = \cos(k) = \cos\left(\frac{2\pi n}{M}\right),
\end{equation}
as shown in Appendix \ref{Appendix:1DEigen}. In these cases, we have
\begin{equation}	
	H_\mathrm{XXZ} 	\left\vert N_k \right\rangle = \frac{M}{4} \Delta \left\vert N_k \right\rangle.
\end{equation}
Moreover, the unimodal extreme sAGP is the highest energy state at the Heisenberg point
$\Delta=1$, and the bimodal extreme sAGP is the ground state for $\Delta=-1$.
The result can also be extended to OBC.
In the interval $-1<\Delta=\cos\left(\frac{2 \pi n}{M}\right)<1$,
the multimodal extreme sAGP are eigenstates of the Hamiltonian, known as scarred states, and they describe non-thermal behavior \cite{regnault2022quantum,melendrez2022real}.

  Reduced density matrices of extreme sAGP states are trivial to compute because all elements are identical (ratios of combinatorial numbers), making it possible to correlate sAGP with low computational cost.  

Multimodal extreme sAGPs turn out to be the lowest energy sAGP states not only for the
XY phase ($-1 \lesssim \Delta \lesssim 1$) of the
1D XXZ model but also for the 2D XXZ and  2D $\mathrm{J_1-J_2}$ models, which will be discussed
in Secs.~\ref{XXZ2D} and \ref{J1J2}.  As with the 1D XXZ model, a multimodal extreme sAGP is the exact ground state in the 2D XXZ Hamiltonian at a specific lattice-dependent value of $\Delta$.

It should be emphasized that not every sAGP is of extreme multimodal form; for example, the sAGP ground state in the 1D XXZ model for $|\Delta| \gtrsim 1$ is usually not extreme sAGP.  We observe, however, that for the spin lattice models that we have studied in this work, the lowest energy sAGP state frequently has multimodal extreme character as obtained from variational optimization.

\subsection{Correlating Spin AGP in the One-Dimensional XXZ Model}

\subsubsection{Incorporating Jastrow-Type Correlators}
\begin{figure}
  \includegraphics[width=\columnwidth]{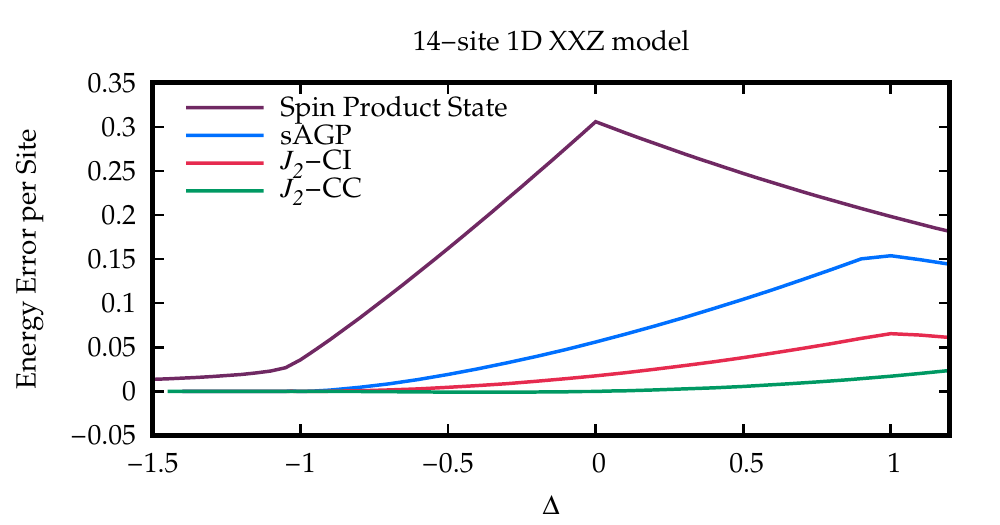}
  \hfill
  \includegraphics[width=\columnwidth]{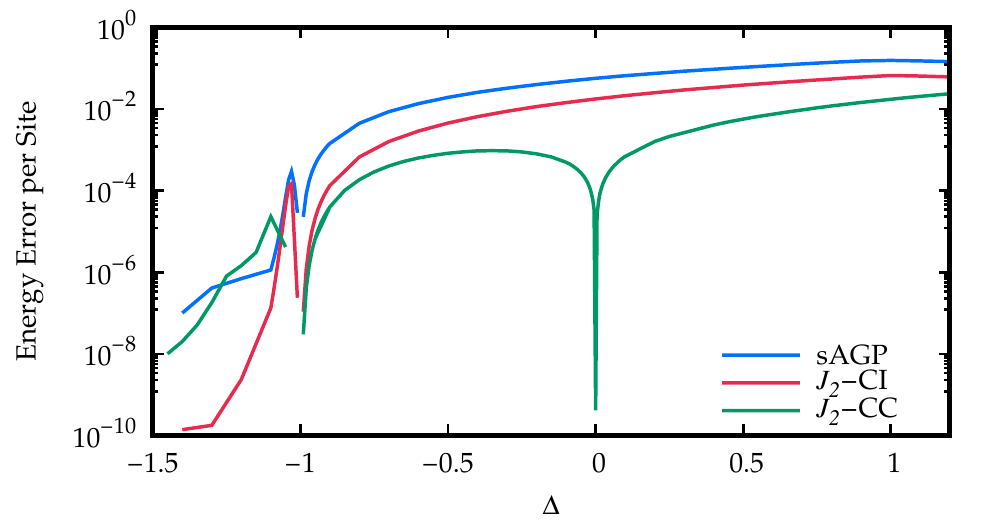}
  \caption{Energy errors for the 14-site 1D XXZ model with open boundary conditions in the $S^z = 0$ sector, on linear scale (top panel) and logarithmic scale (bottom panel).
  The $J_2$-CI and $J_2$-CC methods are based on sAGP. Spin product state results
  are also included in the top panel for comparison.
  \label{Fig:1DXXZJastrow}}
  \end{figure}
  
  \begin{figure}
    \includegraphics[width=\columnwidth]{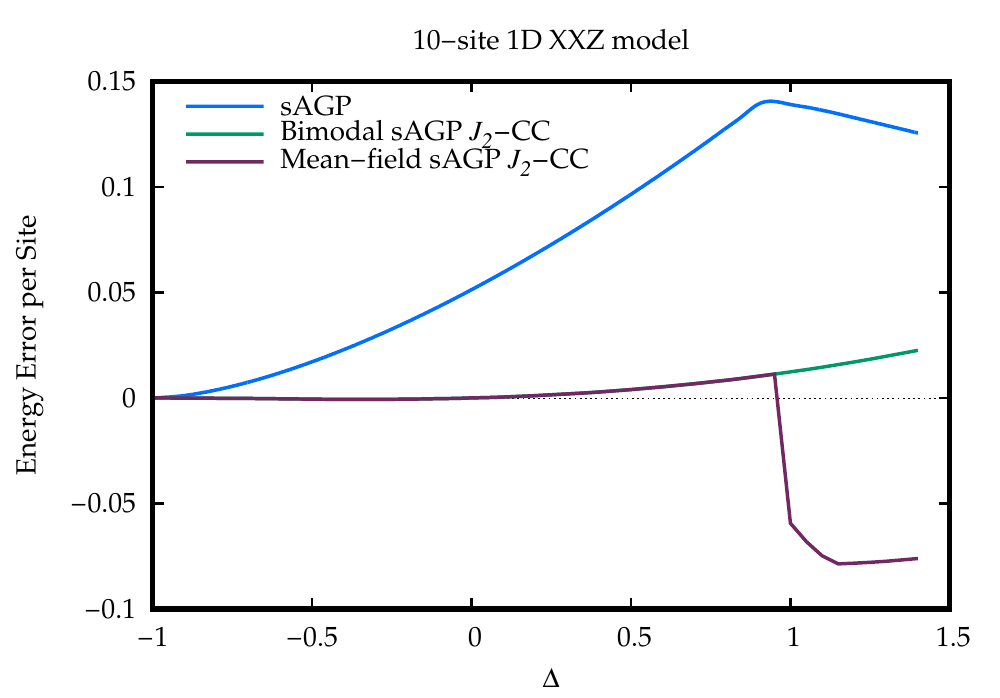}
    \caption{\label{fig J2CC alt}Errors in the sAGP and $J_2$-CC energies based on the mean-field optimized sAGP and the bimodal extreme sAGP in  the 10-site XXZ Heisenberg chain with open boundary conditions in the $S^z=0$ sector.}
  \end{figure}

After studying the properties of sAGP solutions, we can now look at improving them by adding correlation. Correlating AGP with the equivalent of the AGP killing operator presented in previous work \cite{henderson2019geminal},
\begin{align}
K_{pq} &= \eta_p^2 \, P_p^\dagger \, P_q + \eta_q^2 \, P_q^\dagger \, P_p
\\
 &+ \frac{1}{2} \, \eta_p \, \eta_q \, \left(N_p \, N_q - N_p - N_q\right),
\nonumber
\end{align}
is not helpful here. This is because whenever $\eta_p^2 = \eta_q^2$, as in the case of a bimodal extreme sAGP state,
$K_{pq}$ is Hermitian and $K_{pq}^\dag$ also kills sAGP.

\begin{figure}
  \includegraphics[width=\columnwidth]{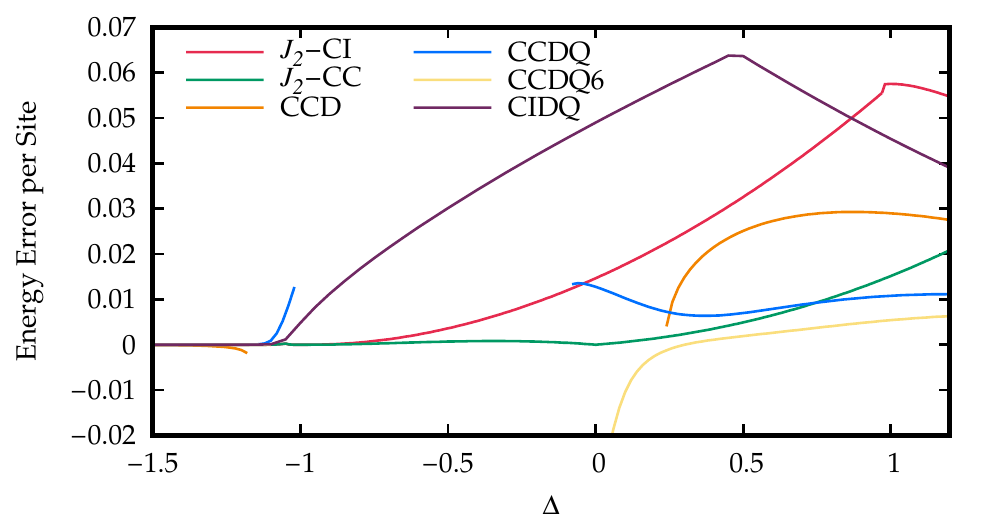}
  \hfill
  \includegraphics[width=\columnwidth]{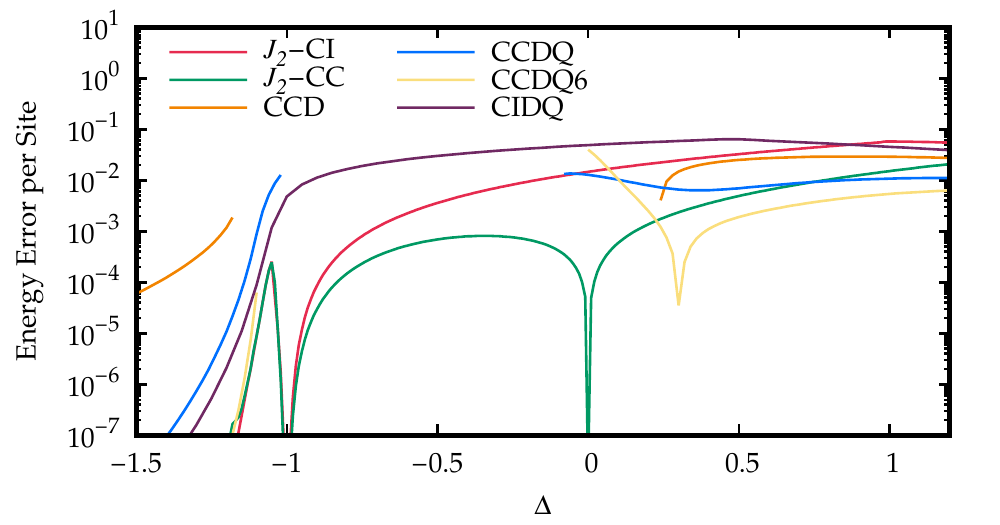}
  \caption{Energy errors for the 12-site 1D antiferromagnetic XXZ model with open boundary conditions in the $S^z = 0$ sector, on linear scale (top panel) and logarithmic scale (bottom panel).
The $J_2$-CI and $J_2$-CC methods are based on sAGP, while CIDQ, CCD, CCDQ, and CCDQ6 are based on spin product state.
  \label{Fig:1DXXZCompare}}
  \end{figure}

Fortunately, we can use Hilbert space Jastrow correlators instead,
which generate the same manifold as do the killing operators in the $\eta_p^2 \neq \eta_q^2$ case
\cite{henderson2020correlating} because both ultimately correspond to a geminal replacement theory \cite{dutta2020geminal}.

\begin{figure*}[t]

  \includegraphics[width=0.375\columnwidth]{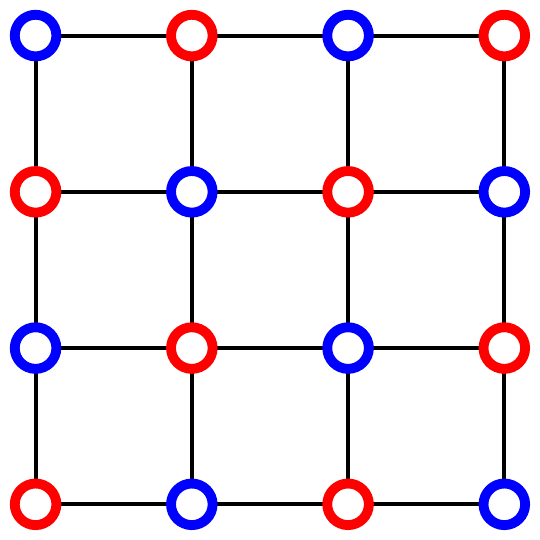}
  \hfill
  \includegraphics[width=0.475\columnwidth]{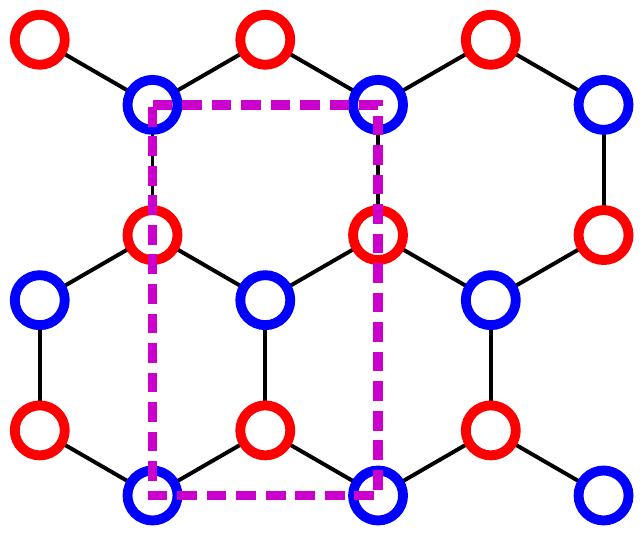}
  \hfill
  \includegraphics[width=0.625\columnwidth]{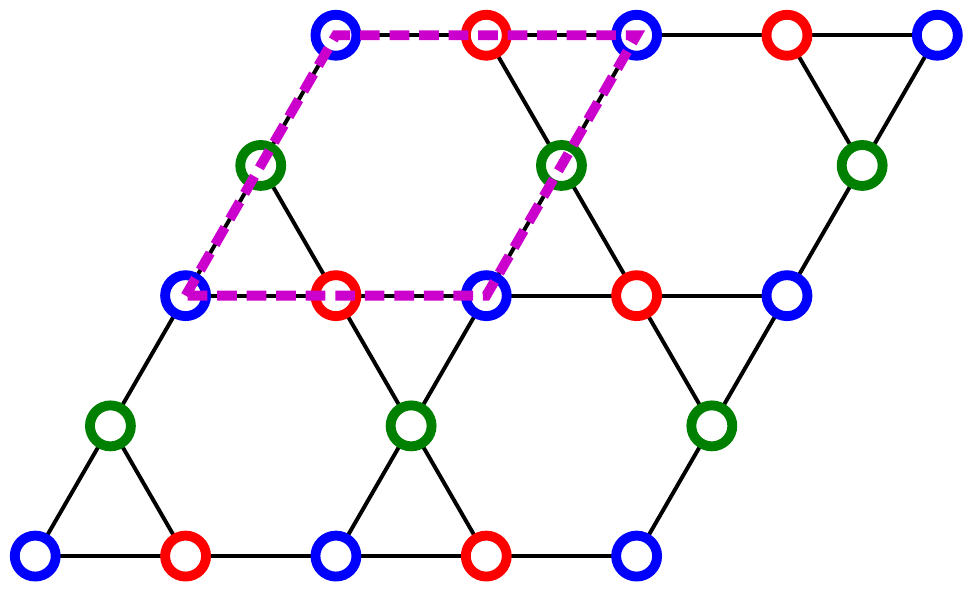}
  \hfill
  \includegraphics[width=0.525\columnwidth]{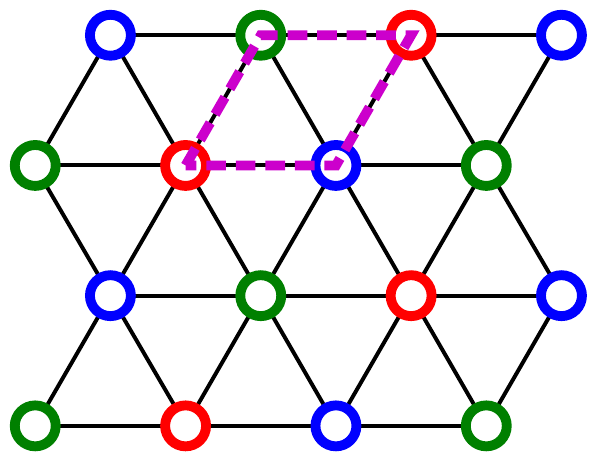}
  \caption{\label{2D lattice}Assorted 2D lattices.  From left to right, these are the square lattice, the honeycomb lattice, the kagome lattice, and the triangular lattice.  The purple dashed shape, wherever present, indicates the smallest rectangular cell for the honeycomb lattice and the unit cell for the kagome lattice and triangular lattice.  The red, blue and green open circles indicate the different $\eta$ values for the sAGP ground state in the XXZ Hamiltonian, which is bimodal extreme for the square and honeycomb lattices, but trimodal extreme for the kagome and triangular lattices.}
\end{figure*}

By Jastrow-type correlators, we mean operators of the form
\begin{subequations}
\begin{align}
J_2 &= \frac{1}{4} \, \sum_{p<q} \alpha_{pq} \, N_p \, N_q
\\
 &\mapsto \sum_{p<q} \alpha_{pq} \, \left(2 \, S_p^z - 1\right) \, \left(2 \, S_q^z - 1\right).
\end{align}
\end{subequations}
Note that we use the italic $J$ to represent the Jastrow operator, which is not to be confused with the roman symbol $\mathrm{J}$ representing Hamiltonian parameters.  Context should make clear which symbol we mean in any event.

Since the lower-order Jastrow operator $J_1 = \sum \alpha_p \, N_p$ already lurks inside $J_2$ \cite{khamoshi2021exploring}, we can define the $J_2$ operator for sAGP as
\begin{equation}
J_2 = \frac{1}{4} \, \sum_{p<q} \alpha_{pq} \, S_p^z \, S_q^z
\end{equation}
and will use this definition hereafter.

The simplest way to correlate sAGP using these operators is by what we refer to as $J_2$-CI, which writes
\begin{equation}
|\textrm{$J_2$-CI}\rangle = J_2 |\mathrm{sAGP}\rangle,
\label{J2CI}
\end{equation}
where we generally use the mean-field optimized sAGP as a reference.  We then evaluate the energy via an expectation value and minimize it with respect to the amplitudes $\alpha_{pq}$. 

Somewhat more sophisticated is $J_2$-CC, where we use an exponential ansatz instead:
\begin{equation}
|\textrm{$J_2$-CC}\rangle = \mathrm{e}^{J_2} |\mathrm{sAGP}\rangle.
\label{J2CC}
\end{equation}
Although intractable in its variational form, a similarity-transformed approach is quite feasible \cite{wahlen2015lie,khamoshi2021exploring}.  The energy and residual equations are
\begin{subequations}
\begin{align}
E_{\textrm{$J_2$-CC}} &= \langle \mathrm{sAGP}| \bar{H} |\mathrm{sAGP}\rangle,
\\
0 &= \langle \mathrm{sAGP}| S_p^z \, S_q^z \, \left(\bar{H} - E_{\textrm{$J_2$-CC}}\right) |\mathrm{sAGP}\rangle,
\label{R_eq_J2CC}
\end{align}
\end{subequations}
where
\begin{equation}
\bar{H} = \mathrm{e}^{-J_2} \, H \, \mathrm{e}^{J_2}.
\end{equation}
Although the commutator expansion of $\bar{H}$ does not truncate,
it can be analytically resummed to yield an expression in terms of exponentials of one-body operators $J_1$,
which act on one sAGP state to produce another \cite{khamoshi2021exploring}. Both  $J_2$-CI and $J_2$-CC have computational costs proportional to $\mathcal{O}(M^4)$ for these lattice models.

Fig.~\ref{Fig:1DXXZJastrow} shows errors of $J_2$-CI and $J_2$-CC for the 14-site antiferromagnetic XXZ model with OBC.  We see that $J_2$-CI eliminates about half the error of sAGP, while the improvement given by $J_2$-CC is significantly larger, with an error one order of magnitude smaller than the error of sAGP itself. This is particularly true when $J_2$-CC is based on the bimodal extreme sAGP everywhere, and not just where this is the lowest energy sAGP (Fig.~\ref{fig J2CC alt}).  A particularly interesting feature is that $J_2$-CC is exact at $\Delta = 0$.  This is true in 1D but not in higher dimensions.  In Appendix \ref{Appendix:BetheAnsatzProof}, we prove this exactness for both open and periodic boundary conditions.

\begin{figure*}
  \includegraphics[width=\columnwidth]{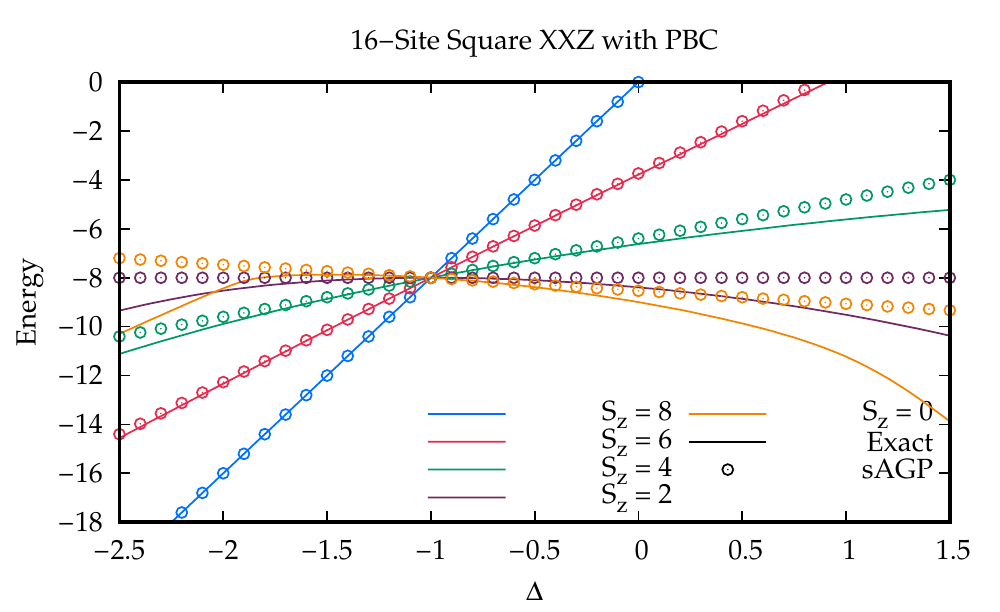}
  \hfill
  \includegraphics[width=\columnwidth]{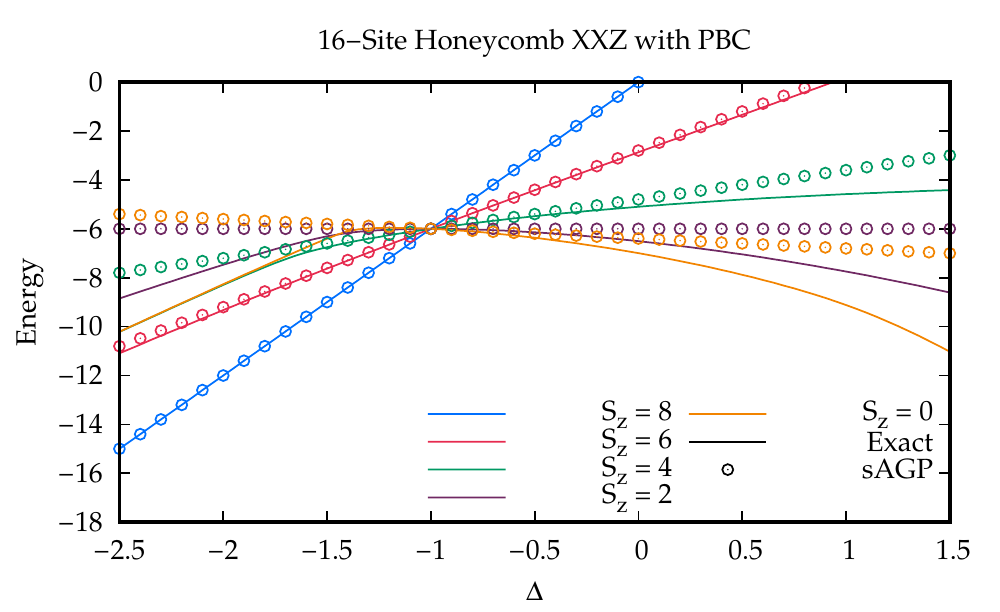}
  \\
  \includegraphics[width=\columnwidth]{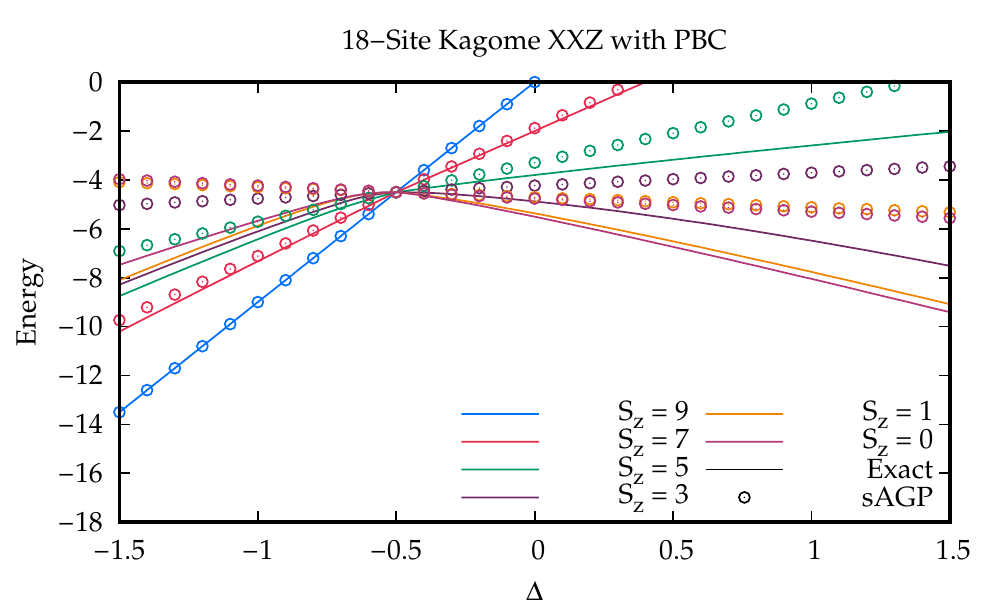}
  \hfill
  \includegraphics[width=\columnwidth]{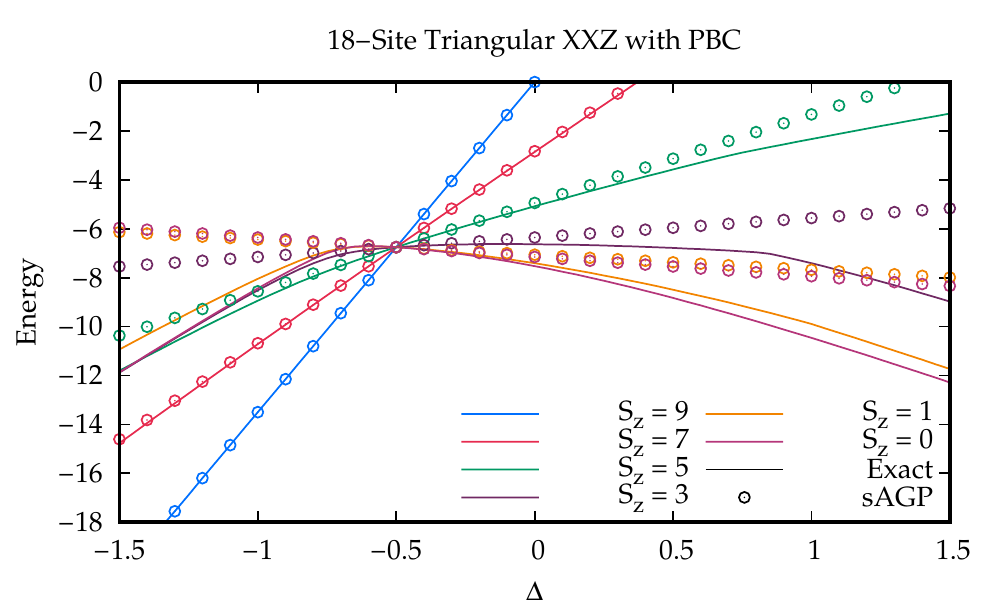}
  \caption{\label{2D multimodal}Multimodal extreme sAGP and exact energies of the XXZ Hamiltonian
  for different lattices and $S^z$ sectors.  The lines correspond to the exact energies and the
  circles to the multimodal extreme sAGP results.  Different colors correspond to different $S^z$ sectors.
  Top left: square lattice.  Top right: honeycomb lattice.  Bottom left: kagome lattice.  Bottom right: triangular lattice. 
  All exact and sAGP results have the same energy for all $S^z$ sectors at $\Delta = -1$ (square or honeycomb lattice) or at $\Delta = -1/2$ (kagome or triangular lattice).
We note sAGP is always exact for $S^z = \pm\frac{M}{2}$ and $S^z = \pm(\frac{M}{2}-1)$ (not shown in the fiugre) for these 2D lattices, for the same reason as in the 1D case discussed in Sec.~\ref{XXZdiffSz}.}
\end{figure*}

\subsubsection{Comparison with Conventional Correlation Methods}

To demonstrate the advantage of sAGP-based correlated methods over conventional correlation methods based
on SPS, we compare their energies for the 12-site antiferromagnetic XXZ model.
Fig.~\ref{Fig:1DXXZCompare} shows the energy errors of sAGP-based $J_2$-CI and $J_2$-CC 
along with SPS-based configuration interaction doubles and quadruples (CIDQ)
and coupled cluster doubles to hextuples (CCD, CCDQ, CCDQ6).
The two sAGP-based correlated methods have the same computational complexity as SPS-based CIDQ and CCDQ,
scaling as $\mathcal{O}(M^4)$, while CCD and CCDQ6 scale as
$\mathcal{O}(M^2)$ and $\mathcal{O}(M^6)$, respectively. 
We note in passing that odd CC excitations (singles, triples, etc.) do not contribute because of $S^z$ symmetry.

The results of sAGP-based methods are generally superior to those of SPS-based methods
with equivalent computational scaling for $\Delta \leq 1$, which corresponds to the ferromagnetic
and XY phases. It is important to note that sAGP is exact
at $\Delta = -1$, whereas conventional coupled cluster calculations break down in this vicinity.
As noted above, $J_2$-CC is also exact at $\Delta = 0$, and it is the most accurate low-scaling correlated
method overall.
One may of course use an $S^z$-broken SPS reference to obtain better CC energies \cite{bishop1996coupled},
but at the cost of breaking physical symmetries of the Hamiltonian, which sAGP and correlated sAGP conserve.

\subsection{The Two-Dimensional XXZ Model}
\label{XXZ2D}

\begin{figure*}
  \includegraphics[width=\columnwidth]{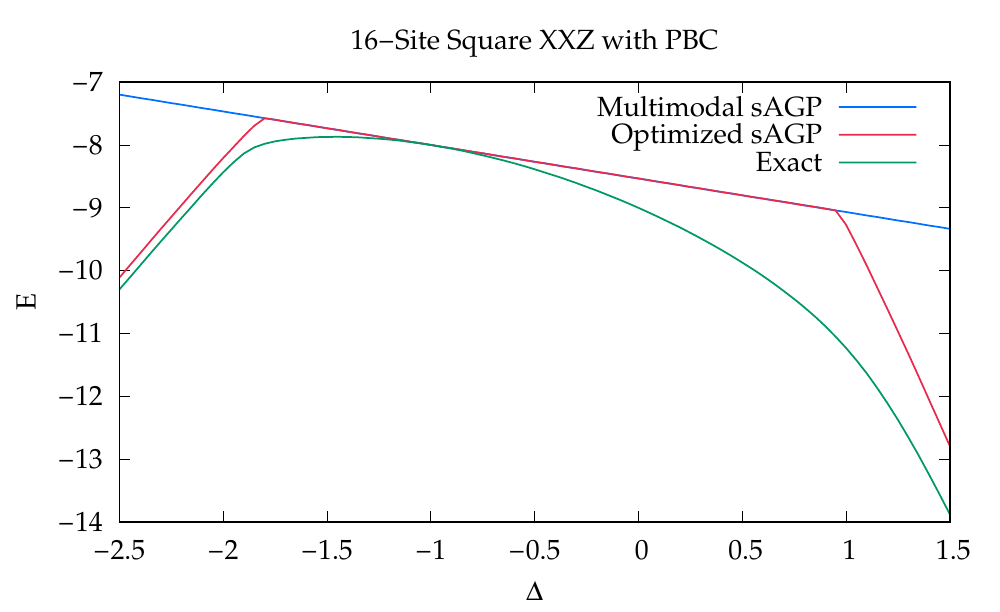}
  \hfill
  \includegraphics[width=\columnwidth]{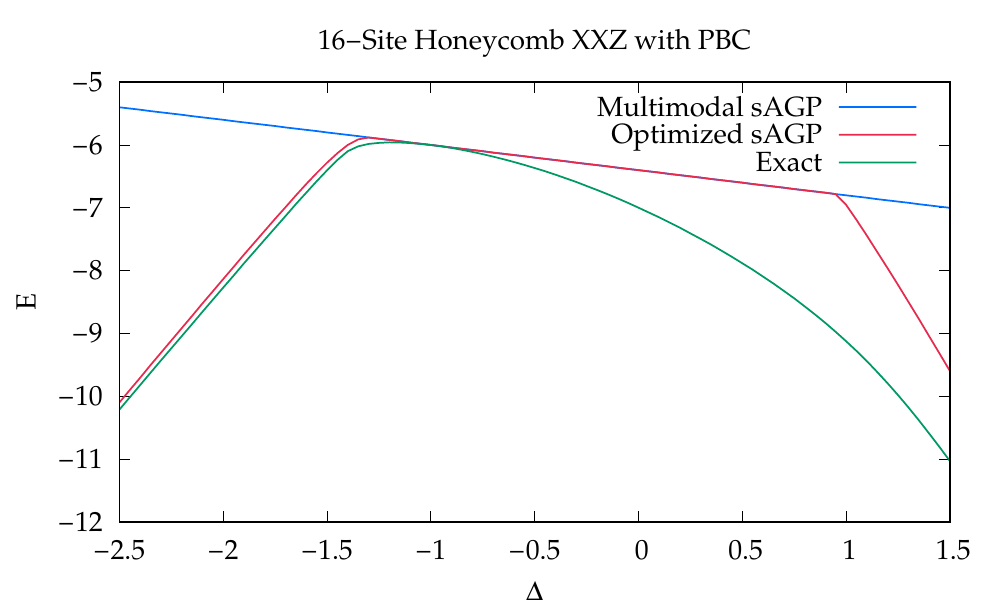}
  \\
  \includegraphics[width=\columnwidth]{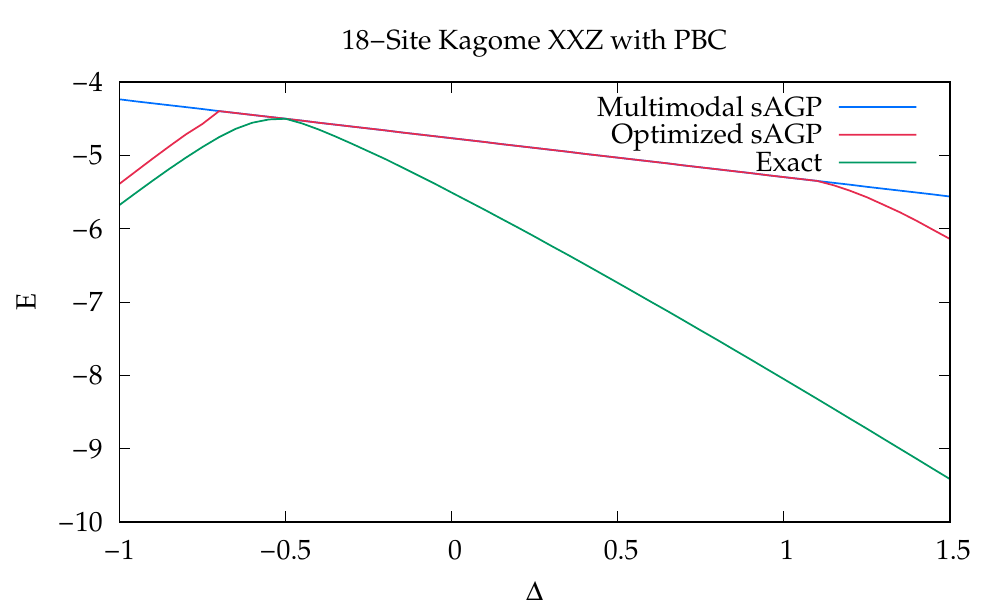}
  \hfill
  \includegraphics[width=\columnwidth]{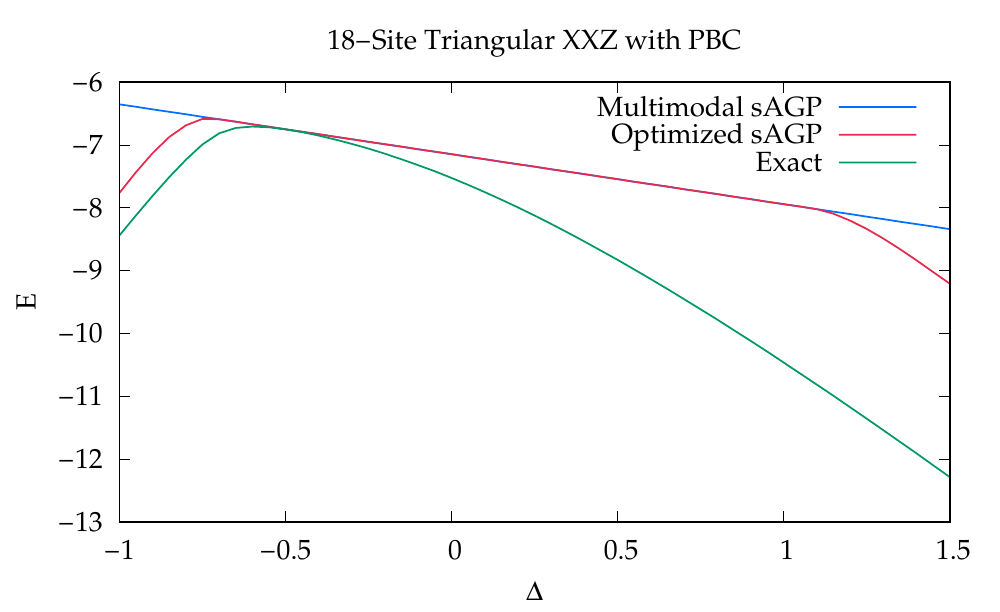}
  \caption{\label{2D multimodal2}Exact energies and those of the multimodal extreme sAGP and mean-field optimized sAGP for the XXZ Hamiltonian with $S^z = 0$.  Top left: 16-site square lattice.  Top right: 16-site honeycomb lattice.  Bottom left: 18-site kagome lattice.  Bottom right: 18-site triangular lattice. }
\end{figure*}

\begin{figure*}
  \includegraphics[width=\columnwidth]{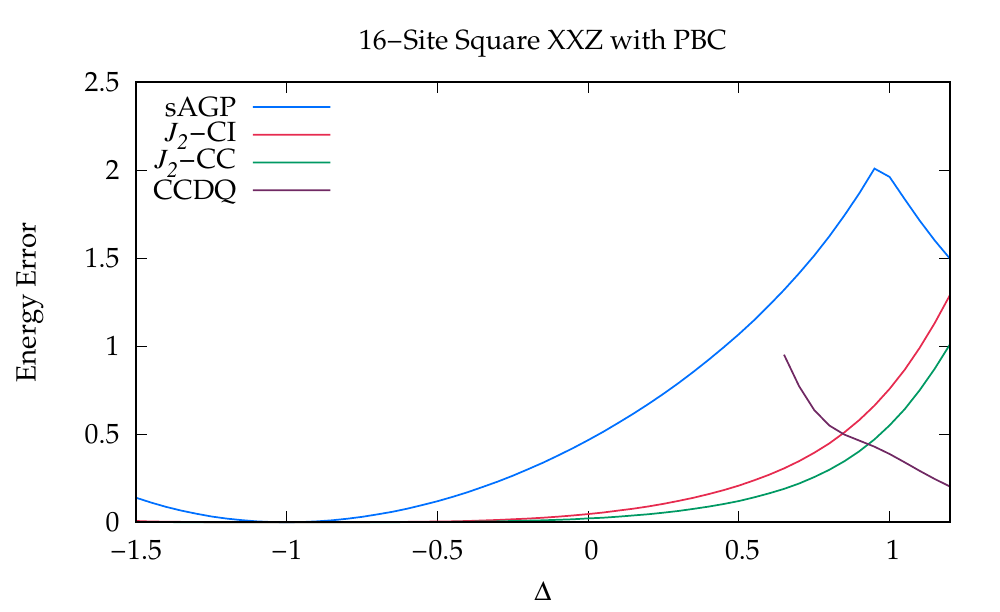}
  \hfill
  \includegraphics[width=\columnwidth]{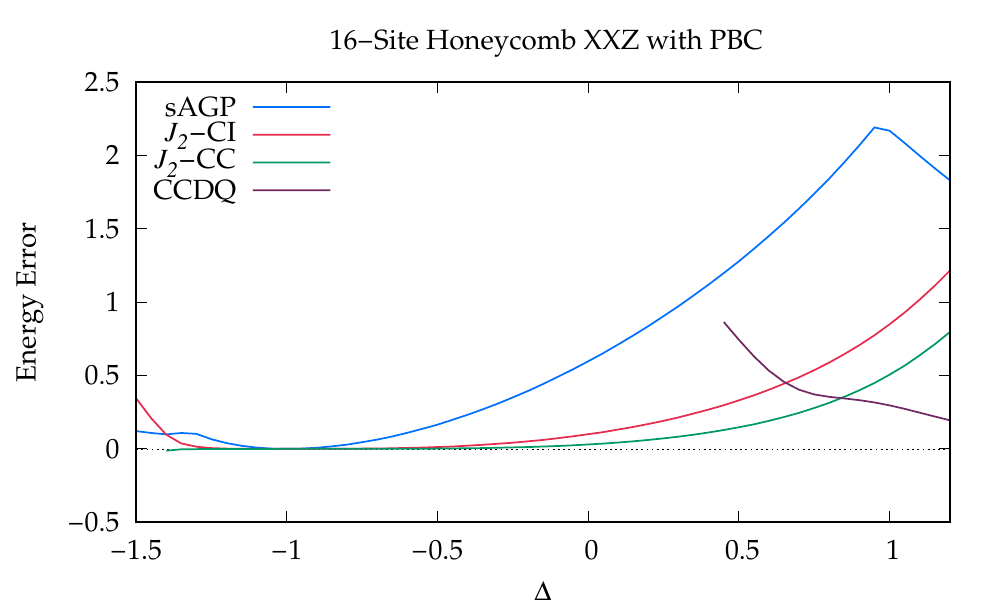}
  \\
  \includegraphics[width=\columnwidth]{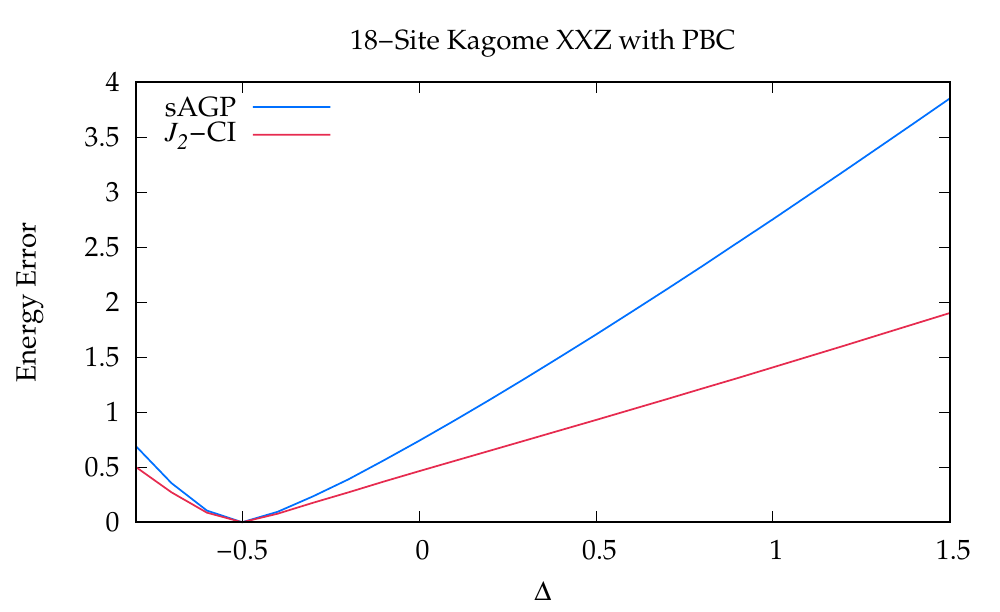}
  \hfill
  \includegraphics[width=\columnwidth]{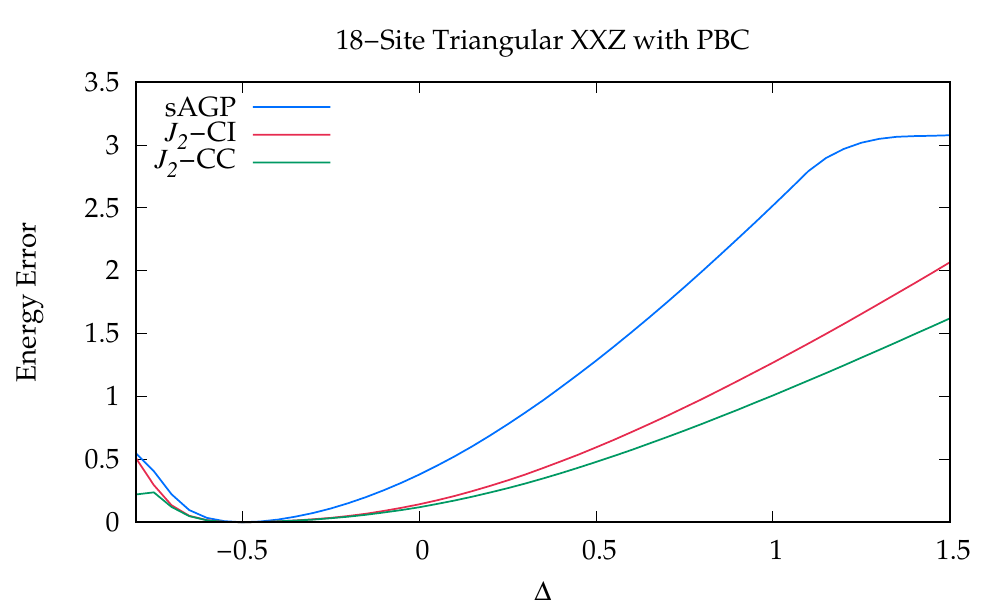}
  \caption{\label{2D-correlated}
  Energy errors for the mean-field optimized sAGP and for $J_2$-CI and $J_2$-CC based on the multimodal extreme sAGP, in various XXZ lattices with $S^z = 0$.  Top left: 16-site square lattice.  Top right: 16-site honeycomb lattice.  Bottom left: 18-site kagome lattice.  Bottom right: 18-site triangular lattice. Conventional CCDQ results are also shown for the square and honeycomb lattice as a comparison. Note CCDQ fails to converge for square and honeycomb lattice at $\Delta<0.65$ and $\Delta<0.45$ respectively. Also Note that $J_2$-CC does not converge for the kagome lattice and has been omitted from the plot. }
\end{figure*}

We next test our methods on several XXZ 2D lattices including the square
lattice, honeycomb lattice, triangular lattice, and kagome lattice
(Fig.~\ref{2D lattice}). 
In Appendix \ref{Appendix:2DGround}, we show analytically that for
both PBC and OBC with certain boundary shapes, the bimodal extreme sAGP is the
ground state of the square and honeycomb lattices at $\Delta = -1$, while the
trimodal ($m=3$) extreme sAGP is the ground state of the triangular and kagome lattices
at $\Delta = - 0.5$. This trimodal extreme sAGP has three distinct $\eta$ values
which we denote by $\eta_1, \eta_2$, and $\eta_3$. As explained in Sec.~\ref{sec:multimodal}, these three distinct 
$\eta$ values are the three cube roots of 1:
\begin{equation}
  \eta_1=1 ,~\eta_2=e^{i\frac{2}{3}\pi},~\eta_3=e^{i\frac{4}{3}\pi.}
\end{equation}
The arrangements of the $\eta$ values in different lattices are illustrated
in Fig.~\ref{2D lattice}.
These analytical results are corroborated by numerical calculations as shown in Fig.~\ref{2D multimodal}.
The ground states of the 2D XXZ models at these special $\Delta$ values have been reported in
Ref.~\cite{pal2021colorful,changlani2018macroscopically,chertkov2021motif},
though they are expressed in a form different from sAGP.

While we do not wish to dwell on these various lattices in detail, we have a few things to point out.

First, as we can see in Fig.~\ref{2D multimodal2}, sAGP is extreme over a range of $\Delta$ for all of the lattices.
As with the 1D case, the sAGP ground state becomes non-extreme around $\Delta = 1$ for all of the 2D lattices considered here.
It also becomes non-extreme for some negative $\Delta$, but the crossover point is lattice-dependent.
We notice that the crossover points for different lattices are correlated with the $\Delta$ values at which the extreme sAGP is exact as discussed above.

Second, as shown in Fig.~\ref{2D-correlated}, $J_2$-CC is no longer exact at $\Delta = 0$ for 2D lattices, as opposed to the 1D case. 
This is reminiscent of Jordan--Wigner transformed Hartree--Fock being exact at $\Delta = 0$
for the 1D spin models but not for their 2D counterparts \cite{doi:10.1063/5.0125124,nishimori2010elements}.
Although the results of $J_2$-CC or $J_2$-CI are not as good in 2D as they are in 1D, they still capture more than half of the correlation energy missing from the mean-field optimized sAGP methods.
They also outperform the conventional SPS-based correlation method (CCDQ here) for $\Delta < 1$. The error of SPS-based CCDQ grows rapidly as $\Delta$ goes below $1$ until it eventually encounters convergence issues. 
Conventional CCDQ fails to converge for triangular and kagome lattices as well.
While $J_2$-CC also has difficulty converging for the kagome lattice, it behaves reasonably well for the triangular lattice.

\begin{figure}
  \includegraphics[width=0.45\columnwidth]{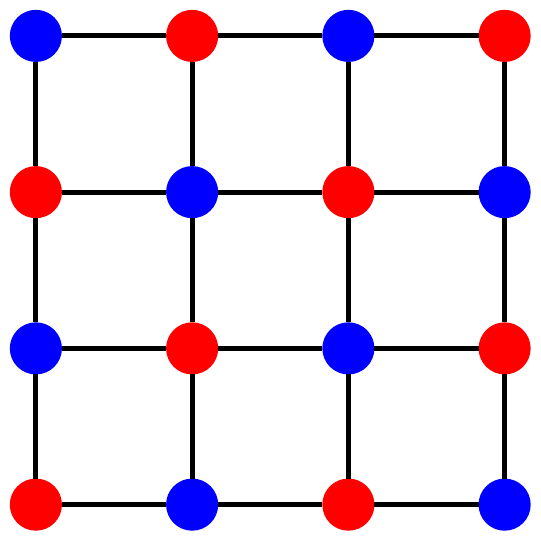}
  \hfill
  \includegraphics[width=0.45\columnwidth]{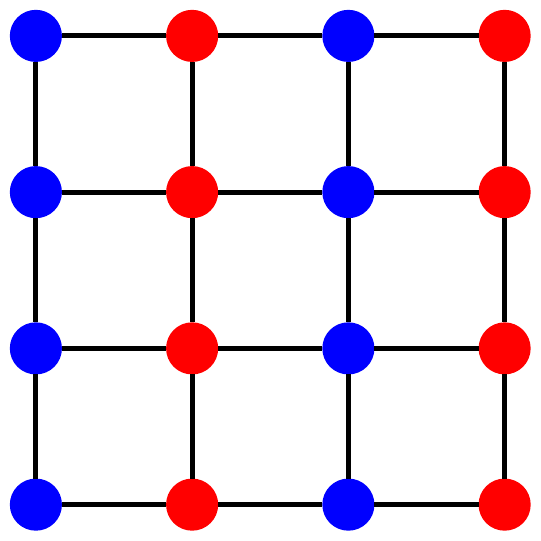}
  \caption{\label{Fig:J1J2NeelColumnar} The sAGP $\eta$ pattern for the $4 \times4$ $\mathrm{J_1-J_2}$ model with PBC. All sites with the same color have the same $\eta$ value. The left figure corresponds to $\mathrm{J_2}<0.5$, and the right $\mathrm{J_2}>0.5$}
\end{figure}

\begin{figure}
  \includegraphics[width=\columnwidth]{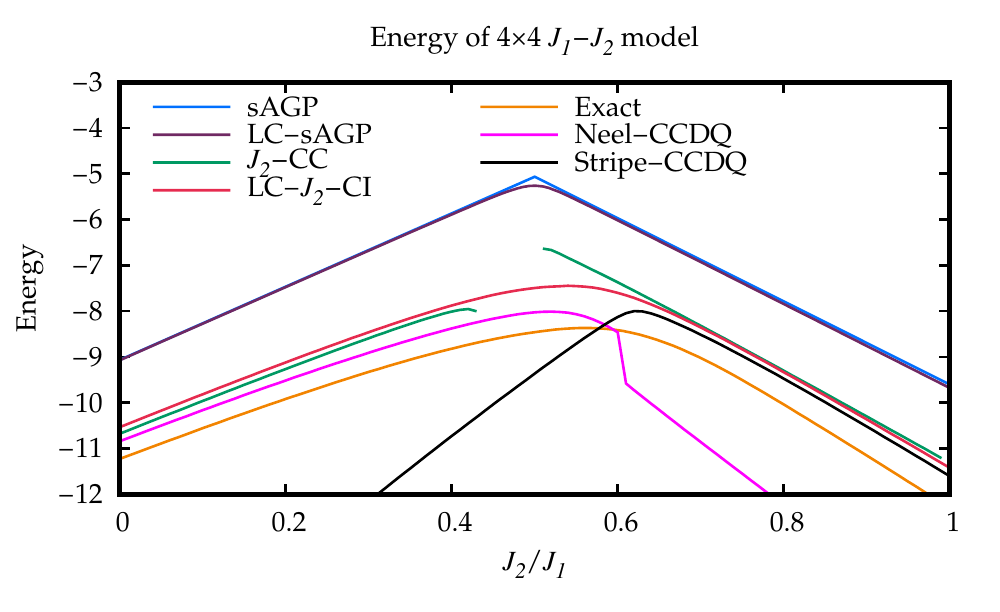}
  \caption{Energy error for the $4\times4$ $\mathrm{J_1-J_2}$ model with PBC. LC-sAGP is a linear combination of 7 bimodal extreme sAGPs. $J_2$-CC and LC-$J_2$-CI are correlated methods based on sAGP and LC-sAGP. N{\'e}el-CCDQ and Stripe-CCDQ are conventional CCDQ results based on different reference states,
    which are included for comparison with sAGP-based methods.
  \label{Fig:$J_1-J_2$}}
  \end{figure}

\subsection{The $\mathrm{J_1-J_2}$ Model}
\label{J1J2}

We also test our sAGP-based methods on the 2D square $\mathrm{J_1-J_2}$ lattice with PBC:
\begin{equation}
  H_\mathrm{J_1-J_2}
  = \mathrm{J_1} \, \sum_{\langle pq \rangle} \left(\vec{S_p}\cdot\vec{S_q}\right)
  + \mathrm{J_2} \, \sum_{\langle\langle pq \rangle\rangle} \left(\vec{S_p}\cdot\vec{S_q}\right),
\end{equation}
where $\langle\langle pq \rangle\rangle$ denotes sites $p$ and $q$ being next-nearest neighbors.
We take $\mathrm{J_1} = 1$, and vary $\mathrm{J_2}$. In the thermodynamic limit, for $\mathrm{J_2} \lesssim 0.45$, the system is in a N{\'e}el order where all spins are antiparallel to their nearest neighbors. And for $\mathrm{J_2} \gtrsim 0.61$, the system is in a well-established striped order with spins parallel in the same column (or row) but antiparallel between columns (or rows) \cite{LIU20221034}. For $\mathrm{J_2} \approx 0.5$, however, the system is in a highly frustrated phase. The ground state is under debate and possible candidates include the plaquette valence-bond state \cite{zhitomirsky1996valence}, the stripe valence-bond state \cite{sachdev1990bond}, or gapless spin liquid state \cite{capriotti2001resonating}.

We find that the optimized sAGP state for the $\mathrm{J_1-J_2}$ model shows a bimodal pattern over
all interaction ranges like the case of XXZ between $-1<\Delta<1$ (Fig.~\ref{Fig:J1J2NeelColumnar}).
For $\mathrm{J_2}<1/2$, the $\eta$ values show a N{\'e}el pattern, while for $\mathrm{J_2}>1/2$, $\eta$ values
exhibit a striped pattern. The two patterns are degenerate at $\mathrm{J_2}=1/2$. As is shown in Table \ref{tab:J1J2 energy}, for small system sizes,
the optimized sAGP is bimodal but non-extreme ($|\eta_1|\neq|\eta_2|$), though the bimodal extreme state ($\eta_1=1, \eta_2=-1$) is still a local minimum. For large system sizes, the bimodal extreme sAGP becomes lower in energy than the non-extreme sAGP.

\begin{table}[h!]
  \caption{Spin AGP energy of the $\mathrm{J_1 - J_2}$ model at $\mathrm{J}_2 = 1/2$ for different system sizes. We see the energy is only dependent on the system size. For small system sizes, the optimized sAGP is bimodal but non-extreme while for large system sizes, the bimodal extreme sAGP becomes lower in energy than the non-extreme sAGP.  \label{tab:J1J2 energy}}

\begin{ruledtabular}
    \begin{tabular}{cccc} 
      \textbf{System Size} & \textbf{Extreme} & \textbf{Non-extreme}& \textbf{Energy Difference}\\
      \hline
      $4\times4$ & -5.0667 & -5.2672 & 0.2005\\
      $4\times8$ &-9.0323 & -9.2417 & 0.2094 \\
      $4\times16$ & -17.0245   & -17.2296  & 0.2050\\
      $8\times8$ & -17.0245   & -17.2296  & 0.2050\\
      $8\times12$ & -25.1109      & -25.2256 & 0.1146\\
      $12\times12$ & -37.4387 & -37.2230 & -0.2157\\
      $16\times16$ & -66.4843 & -65.2206 & -1.2637\\
    \end{tabular}
\end{ruledtabular}
\end{table}

Fig.~\ref{Fig:$J_1-J_2$} shows the energies of the bimodal extreme sAGP and its correlated methods
for the $4\times4$ $\mathrm{J_1-J_2}$ model. The two branches of the sAGP curve correspond to the two bimodal extreme patterns (N{\'e}el versus striped).

The $J_2$-CC (Eqn.~\eqref{J2CC}) energy exhibits a discontinuity at $\mathrm{J_2} = 1/2$
because of the two branches of the reference sAGP.
Moreover, for $0.43<\mathrm{J_2}<0.5$ (the tail of the left branch in Fig.~\ref{Fig:$J_1-J_2$}),
the $J_2$-CC residual equations fail to converge.

\begin{figure}
  \includegraphics[width=0.45\columnwidth]{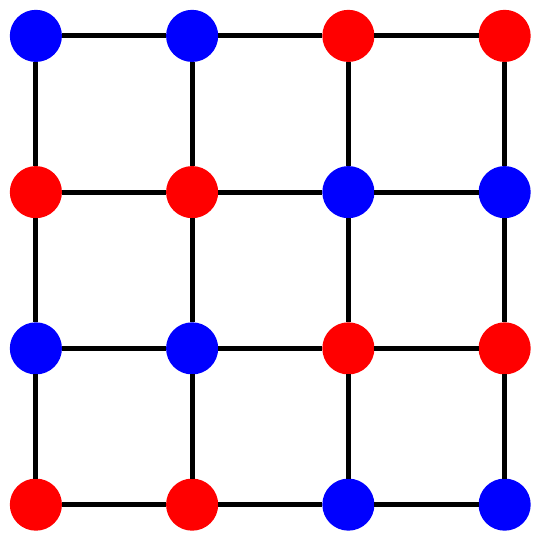}
  \hfill
  \includegraphics[width=0.45\columnwidth]{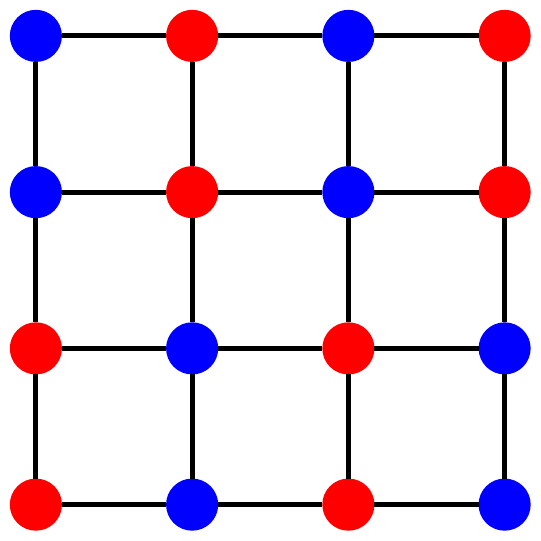}
  \includegraphics[width=0.45\columnwidth]{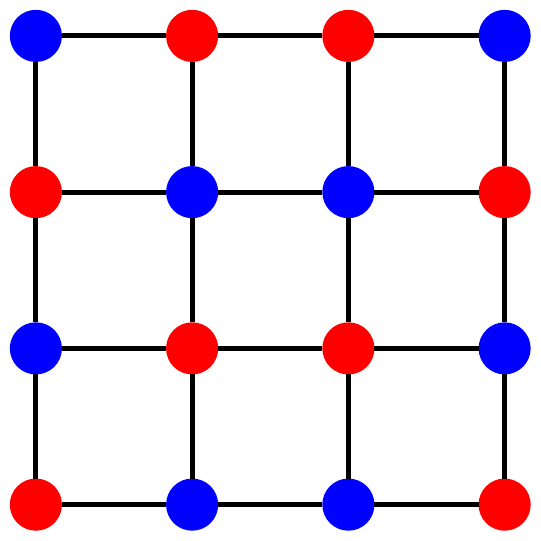}
  \hfill
  \includegraphics[width=0.45\columnwidth]{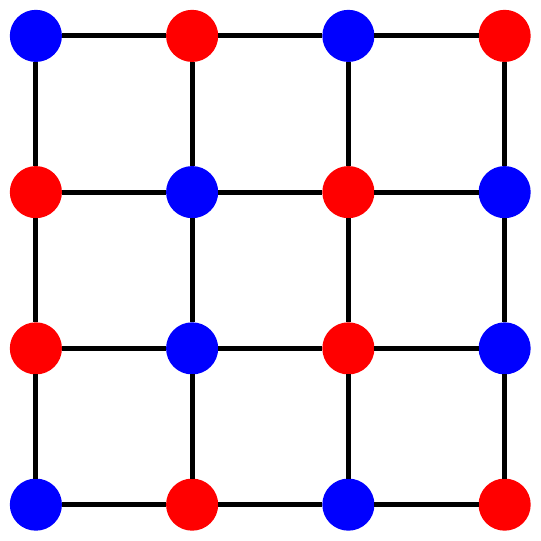}
  \caption{\label{Fig:J1J2Int}The four intermediate bimodal extreme sAGP states necessary for LC-sAGP and LC-$J_2$-CI for the $4\times4$ $\mathrm{J_1-J_2}$ model with PBC.}
\end{figure}

In order to remove the discontinuity and produce well-behaved curves, we consider a reference state that is a linear combination of the relevant sAGPs (LC-sAGP).
This is simply an sAGP-based non-orthogonal CI \cite{dutta2021construction}.
We find that at least 7 bimodal extreme sAGPs are needed if we want to approximate the exact ground state
(with additional $J_2$-CI-type correlation; vide infra).
They include the bimodal extreme sAGP with the N{\'e}el pattern and those with the column-wise and row-wise
striped patterns, as well as four additional intermediate bimodal extreme sAGP states shown in
Fig.~\ref{Fig:J1J2Int}. These intermediate bimodal extreme sAGPs exhibit a pattern between
N{\'e}el and striped where each site has only one nearest neighbor that shares its $\eta$ value.

We see that the LC-AGP is well-behaved near $\mathrm{J_2} = 1/2$ but offers little quantitative improvement over a single sAGP elsewhere.  In practice, this means that $J_2$-CC or $J_2$-CI based on this LC-AGP looks little different from the corresponding methods based on the mean-field optimized sAGP, except for $\mathrm{J_2} \approx 1/2$.  Thus, we consider linear combinations of $J_2$-CI states as well, shown in Fig.~\ref{Fig:$J_1-J_2$} as LC-$J_2$-CI.  This LC-$J_2$-CI is roughly parallel to the exact result, and is comparable to $J_2$-CC, but is correctly smooth everywhere.

For comparison, conventional CCDQ was also implemented for the $\mathrm{J_1 - J_2}$ model with the N{\'e}el and striped SPS as the reference state, denoted as N{\'e}el-CCDQ and Stripe-CCDQ, respectively, in Fig.~\ref{Fig:$J_1-J_2$}. 
In this case, $J_2$-CC and conventional CCDQ 
are of roughly similar quality. Both behave poorly in the frustrated region $J_2 \approx 1/2$.  
One great advantage of Jastrow-type correlators
over conventional particle--hole-type correlators is
that the former, as a similarity transformation,
can be solved over any reference state.  Future work will explore the use of these  $J_2$ correlators on linear combinations of AGPs which go beyond the simple extreme bimodal AGPs used in Fig 11.

\section{Conclusions}

In this article, we have studied sAGP and several sAGP-based correlation methods for the 1D and 2D XXZ models, and the 2D $\mathrm{J_1-J_2}$ model.
With our $\mathcal{O}(M^3)$ implementation of mean-field optimized sAGP, we find that optimized sAGP can capture the phase transitions of the XXZ Heisenberg chain and 2D lattices. Furthermore, we show that the optimized sAGP states turn out to be multimodal extreme for the $\mathrm{J_1-J_2}$  model and the XY phase of the XXZ model, reflecting the translational symmetry of these states. The fact that all $\eta$ have the same absolute value makes the calculation of correlation methods based
on sAGP even easier. These facts suggest that sAGP should be a good reference state for these spin systems.

Though correlation methods based on killing operators \cite{henderson2019geminal} are not feasible for sAGP, we show that Jastrow operators can serve as good correlators for spin systems. Both $J_2$-CI and $J_2$-CC yield a significant improvement over mean-field optimized sAGP with reasonable computational cost; $J_2$-CC behaves especially well in the XY phase $-1\lesssim\Delta \lesssim 1$ for the XXZ chain, and is exact at $\Delta=0$ in 1D.

We have also shown that for the 2D $\mathrm{J_1-J_2}$  model, there are multiple important bimodal extreme sAGP states. The LC-sAGP approach uses a linear combination of these important sAGP states and makes the transition between the N{\'e}el pattern and striped pattern smooth. The LC-$J_2$-CI energy result on $\mathrm{J_1-J_2}$  model is almost parallel to the exact one.

Thus far, we have considered only energies. The behavior of our techniques for correlation functions and other properties will be reported in future work.

\begin{acknowledgments}
This work was supported by the U.S. National Science Foundation under Grant No.~CHE-2153820. G.E.S. is a Welch Foundation Chair (Grant No.~C-0036). J.D. acknowledges financial support from Projects No.~PGC2018-094180-B-I00 and PID2022-136992NB-I00 (MCIU/AEI/FEDER, EU).
\end{acknowledgments}

\appendix
\section{Exactness of $J_2$-CC for 1D XXZ at $\Delta = 0$
\label{Appendix:BetheAnsatzProof}}

A general wave function for an $M$-site 1D spin-$\frac{1}{2}$ system can be written as
\begin{equation}
  |\psi\rangle
  = \sum_{1 \leq p_1 < \ldots p_N \leq M}
  \psi(p_1, \ldots,p_N)S_{p_1}^{\dagger} \ldots S_{p_N}^{\dagger} \left\vert \Downarrow  \right\rangle 
\end{equation}
where $\psi(p_1, \ldots,p_N)$ is the amplitude for the $N$ $\uparrow$-spins at sites $p_1, \ldots,p_N$. 

Exact eigenvalues and eigenstates of the 1D XXZ model with periodic boundary conditions can be found by the Bethe ansatz, where the ground state amplitude can be written as
\begin{equation}
\label{eqn:1}
\psi(p_1, \ldots p_N) = \sum_{\sigma\in S_N}A(\sigma) \text{exp} \left(i\sum_{j=1}^{N} k_{\sigma(j)}p_j\right).
\end{equation}
The summation runs over all $N!$ permutations of $1, \ldots , N$. The amplitudes $A$ relate to the scattering
matrix $S$ through
\begin{equation}
  A(\nu) = S(k_i, k_j) A(\sigma),
\end{equation}
where the permutation $\nu$ is related to the permutation $\sigma$ by swapping $i$ with $j$, and
\begin{equation}
    S(k_i, k_j) = -\frac{e^{i(k_i+k_j)} - 2\Delta e^{ik_j} + 1}{e^{i(k_i+k_j)} - 2\Delta e^{ik_i} + 1}.
\end{equation}

For the case $\Delta = 0$, $S(k_i, k_j) = -1$, and we can choose $A(\sigma) = (-1)^{\text{sgn}(\sigma)}$. The parameters $k_1,\ldots,k_n$ in Eqn.~\eqref{eqn:1} can be solved by the Bethe ansatz equations:
\begin{equation}
    e^{ik_iM} = \prod_{j\neq i}S(k_i, k_j).
\end{equation}
For even $N$, the equations reduce to
\begin{equation}
    e^{ik_iM} = -1
\end{equation}
and $k_i$ are
\begin{equation}
  k_i - \pi = \{ \pm \pi / M, \pm 3\pi / M, \pm 5\pi / M \ldots \}.
\end{equation}
The amplitude $\psi(p_1,\ldots,p_N)$ can thus be written as 
\begin{equation}
\label{eqn:2}
  \psi(p_1, \ldots,p_N) = \det(C)
\end{equation}
where the matrix $C$ is defined by $C_{ij} = e^{ik_ip_j}$ and can be recognized as a Vandermonde matrix. Therefore,
\begin{equation}
\label{eqn:3}
\det(C)
= \prod_{1 \leq i<j \leq N}\sin\big(\frac{\pi}{M} (p_j-p_i) \big) \prod_{l=1}^N e^{i\pi p_l}.
\end{equation}

According to Eqns.~\eqref{eqn:2} and \eqref{eqn:3}, the wave function can be written as $J_2$-CC on the bimodal extreme sAGP:
\begin{equation}
    |\psi\rangle = e^{J_2}|\mathrm{sAGP}\rangle
\end{equation}
with $\eta_p = e^{i\pi p}$, and $J_2$ coefficients satisfying
\begin{equation}
    \alpha_{pq} = \ln\left(\sin\big(\frac{\pi}{M} (q-p)\big)\right)
\end{equation}
for all $1 \leq p<q \leq M$.

For open boundary conditions, the derivation is essentially the same, and the ground state amplitude can still be written as a determinant, but now
\begin{widetext}
\begin{equation}
    \text{det}(C) = \prod_{1 \leq i<j \leq N}2\left(\text{cos}\left(\frac{\pi p_j}{M+1}\right) - \text{cos}\left(\frac{\pi p_i}{M+1}\right)\right)  \prod_{l=1}^{N} \sin\left( \frac{\pi p_l}{M+1} \right).
\end{equation}
\end{widetext}
This can be written as $J_2$-CC on sAGP with coefficients
\begin{subequations}
\begin{align}
  \eta_{p}
  &= \sin\left( \frac{\pi p}{M+1} \right),\\
  \alpha_{pq}
  &= \ln\left(2\left(\text{cos}\left(\frac{\pi q}{M+1}\right) - \text{cos}\left(\frac{\pi p}{M+1}\right)\right)\right).
\end{align}
\end{subequations}

These $\eta$ values are not extreme. However, since the $J_2$ operator contains $J_1$, and $J_1$ transforms the  $\eta$ values \cite{khamoshi2021exploring}, this means $J_2$-CC on bimodal extreme sAGP is also exact.

\section{Multimodal Extreme sAGP as the Eigenstate of 1D XXZ with PBC}
\label{Appendix:1DEigen}
We want to show the multimodal extreme sAGP Eqn.~(\ref{N}) generated by the $K^{+}_k$ operator Eqn.~(\ref{Koperators}) is an eigenstate of the 1D XXZ Hamiltonian with PBC when $\Delta=\cos k$.

First we compute the commutators of $H_\mathrm{XXZ}$ with $K^{+}_k$.  Using
\begin{widetext}
\begin{subequations}
\begin{align}
  &\left[\sum_{p=1}^M \frac{1}{2}  \left(S_p^+ \, S_{p+1}^- + S_p^- \, S_{p+1}^+\right), K^+_k\right]=-\sum_{p=1}^M\left(e^{ik}e^{ikp}S_p^+S_{p+1}^z+e^{ikp}S_{p+1}^+S_p^z\right),
  \\
  &\left[\sum_{p=1}^M S_p^z S_{p+1}^z, K^+_k\right]=\sum_{p=1}^M\left(e^{ik}e^{ikp}S_{p+1}^+S_{p}^z+e^{ikp}S_{p}^+S_{p+1}^z\right),
\end{align}
\end{subequations}
we obtain
\begin{align}
\left[H_\mathrm{XXZ}, K^+_k\right]&=\left(\Delta-e^{ik}\right)\sum_{p=1}^M e^{ikp}S_{p}^+S_{p+1}^z+\left(\Delta e^{ik}-1\right)\sum_{p=1}^M e^{ikp}S_{p+1}^+S_{p}^z,
\end{align}
\begin{subequations}
\begin{align}
  \left[\left[H_\mathrm{XXZ}, K^+_k\right],K^+_k\right]&=\left(\Delta-e^{ik}\right)e^{ik}\sum_{p=1}^M e^{2ikp}S_{p}^+S_{p+1}^+ +\left(\Delta e^{ik}-1\right)\sum_{p=1}^M e^{2ikp}S_{p+1}^+S_{p}^+ \\
  &=\left(2\Delta e^{ik}-e^{2ik}-1\right)\sum_{p=1}^M e^{2ikp}S_{p}^+S_{p+1}^+.
\end{align}
\end{subequations}
We also have
\begin{subequations}
\begin{align}
  \left[H_\mathrm{XXZ}, K^+_k\right] \ket{\Downarrow}
  &= -\frac{1}{2}\left(\left(\Delta-e^{ik}\right)\sum_{p=1}^M e^{ikp}S_{p}^++\left(\Delta e^{ik}-1\right)\sum_{p=1}^M e^{ikp}S_{p+1}^+\right)\ket{\Downarrow}\\ 
  &= -\frac{1}{2}\left(\left(\Delta-e^{ik}\right)\sum_{p=1}^M e^{ikp}S_{p}^++\left(\Delta e^{ik}-1\right)e^{-ik}\sum_{p=1}^M e^{ik(p+1)}S_{p+1}^+\right)\ket{\Downarrow}\\ 
  &= -\frac{1}{2}\left(2\Delta-e^{ik}-e^{-ik}\right)\sum_{p=1}^M e^{ikp}S_{p}^+\ket{\Downarrow}.
\end{align}
\end{subequations}
When $\Delta =\cos k$, we have $\left(2\Delta -e^{ik}-e^{-ik}\right)=0$, thus
\begin{subequations}
\begin{align} 
  \left[\left[H_\mathrm{XXZ}, K^+_k\right],K^+_k\right]=0,\\
  \left[H_\mathrm{XXZ}, K^+_k\right] \ket{\Downarrow}=0.
\end{align}
\end{subequations}

Then we can calculate $H_\mathrm{XXZ} 	\left\vert N_k\right\rangle$:
\begin{subequations}
\begin{align}
  &H_\mathrm{XXZ} 	\left\vert N_k\right\rangle= H_\mathrm{XXZ}~(K^+_k)^N	\ket{\Downarrow}\\
  &=N(K^+_k)^{N-1}\left[H_\mathrm{XXZ}, K^+_k\right] \ket{\Downarrow}+\frac{N(N-1)}{2}(K^+_k)^{N-2}\left[\left[H_\mathrm{XXZ}, K^+_k\right],K^+_k\right] \ket{\Downarrow}+(K^+_k)^N H_\mathrm{XXZ} \ket{\Downarrow}\\
  &=\frac{M}{4}\Delta (K^+_k)^N	\ket{\Downarrow} \\
  &= \frac{M}{4}\Delta \left\vert N_k\right\rangle.
\end{align}
\end{subequations}
\end{widetext}
We see that the multimodal extreme sAGP $\left\vert N_k\right\rangle$ becomes an eigenstate of the Hamiltonian $H_\mathrm{XXZ}$ in 1D, with PBC.

\section{Multimodal Extreme sAGP as the Ground State of Colorable XXZ for Certain $\Delta$}
\label{Appendix:2DGround}
The proof in the previous appendix relies on properties of the $K^+_k$ operator to show that extreme multimodal sAGP is an eigenstate of the 1D XXZ Hamiltonian with PBC.  In fact, as we have noted in the text, multimodal extreme sAGP is the exact ground state at certain values of $\Delta$ even in multiple dimensions. Here, we sketch a proof of this claim.

\subsection{Bimodal Extreme sAGP for Bipartite Lattices}
Bimodal extreme sAGP is the ground state for the 1D XXZ chain and 2D square and honeycomb lattices at $\Delta = -1$. In fact, it is the ground state at this $\Delta$ for any lattice such that all neighboring sites are of different colors (i.e. for any bipartite lattice).

Say $p,q$ are neighboring sites. Let 
\begin{equation}
  H_{pq}=\frac{(S_{p}^+ S_{q}^-+S_{p}^- S_{q}^+)}{2}+\Delta S_p^z S_q^z.
  \label{singlebond}
\end{equation}
The XXZ Hamiltonian can then be written as
\begin{equation}
   H_{\mathrm{XXZ}} =\sum_{\langle pq \rangle}H_{pq}.
\end{equation}
We will show that bimodal extreme sAGP is the ground state not only of the whole Hamiltonian
$H_{\mathrm{XXZ}}$, but also for each bond $H_{pq}$.

The sAGP is 
\begin{subequations}
\begin{align}
  \ket{\mathrm{sAGP}}
  &=\frac{1}{N!}(\sum_{p=1}^{M}\eta_{p}S_p^+)\ket{\Downarrow}
\\
  &=\sum_{1\leq p_1<\cdots<p_N\leq M}\eta_{p1}\cdots\eta_{p_N} S_{p1}^+ \cdots S_{p_N}^+\ket{\Downarrow}.
\end{align}
\end{subequations}
For a given pair of nearest neighbors $p$ and $q$, sAGP can be written as
\begin{align}
&\ket{\mathrm{sAGP}}
\\
&\quad =\sum_{\sim}c_{\uparrow\uparrow}(\sim)\eta_p\eta_q\ket{\sim\uparrow_p\uparrow_q\sim}+\sum_{\sim}c_{\downarrow\downarrow}(\sim)\ket{\sim\downarrow_p\downarrow_q\sim}
\nonumber
\\
&\quad+\sum_{\sim}c_{\uparrow\downarrow}(\sim)\eta_p\ket{\sim\uparrow_p\downarrow_q\sim}+\sum_{\sim}c_{\downarrow\uparrow}(\sim)\eta_q\ket{\sim\downarrow_p\uparrow_q\sim}
\nonumber.
\end{align}
Here, $\sim$ represents all possible situations of the sites other than $p$ and $q$. $c_{\uparrow\uparrow}(\sim)$, $c_{\downarrow\uparrow}(\sim)$, $c_{\uparrow\downarrow}(\sim)$, $c_{\downarrow\downarrow}(\sim)$ are the products of the $\eta$ values of spin-$\uparrow$ sites in each respective $\sim$.

The two summations for $\ket{\sim\downarrow_p\uparrow_q\sim}$ and $\ket{\sim\uparrow_p\downarrow_q\sim}$ are the same, as there are $M-2$ other sites, $N-1$ of which have $\uparrow$ spin. For the same reason $c_{\uparrow\downarrow}(\sim)=c_{\downarrow\uparrow}(\sim)$, so
\begin{align}
\ket{\mathrm{sAGP}}
  &=\sum_{\sim}c_{\uparrow\uparrow}(\sim)\eta_p\eta_q\ket{\sim\uparrow_p\uparrow_q\sim}+\sum_{\sim}c_{\downarrow\downarrow}(\sim)\ket{\sim\downarrow_p\downarrow_q\sim}
\nonumber
\\
  &+\sum_{\sim}c_{\uparrow\downarrow}(\sim)(\eta_p\ket{\sim\uparrow_p\downarrow_q\sim}+\eta_q\ket{\sim\downarrow_p\uparrow_q\sim}).
\end{align}
For bimodal extreme sAGP, $\eta_p=-\eta_q$, so that
\begin{align}
\ket{\mathrm{sAGP}}
  &=-\sum_{\sim}c_{\uparrow\uparrow}(\sim)\ket{\sim\uparrow_p\uparrow_q\sim}+\sum_{\sim}c_{\downarrow\downarrow}(\sim)\ket{\sim\downarrow_p\downarrow_q\sim}
\nonumber
\\
  &+\sum_{\sim}c_{\uparrow\downarrow}(\sim)\eta_p(\ket{\sim\uparrow_p\downarrow_q\sim}-\ket{\sim\downarrow_p\uparrow_q\sim}).
\end{align}

Now note that
\begin{align}
&H_{pq}\ket{\sim\uparrow_p\uparrow_q\sim}
 =\Delta S_p^z S_q^z\ket{\sim\uparrow_p\uparrow_q\sim}=\frac{\Delta}{4}\ket{\sim\uparrow_p\uparrow_q\sim}
\\
&H_{pq}\ket{\sim\downarrow_p\downarrow_q\sim}
 =\Delta S_p^z S_q^z\ket{\sim\downarrow_p\downarrow_q\sim}=\frac{\Delta}{4}\ket{\sim\downarrow_p\downarrow_q\sim}
\end{align}
\begin{subequations}
\begin{align}
  &H_{pq}(\ket{\sim\uparrow_p\downarrow_q\sim}-\ket{\sim\downarrow_p\uparrow_q\sim})
\\
  =&(\frac{(S_{p}^+ S_{q}^-+S_{p}^- S_{q}^+)}{2}+\Delta S_p^z S_q^z)(\ket{\sim\uparrow_p\downarrow_q\sim}-\ket{\sim\downarrow_p\uparrow_q\sim})
\\
  =&(-\frac{1}{2}-\frac{1}{4}\Delta)(\ket{\sim\uparrow_p\downarrow_q\sim}-\ket{\sim\downarrow_p\uparrow_q\sim}).
\end{align}
\end{subequations}
When $\Delta =-1$, we obtain
\begin{equation}
 H_{pq}\ket{\mathrm{sAGP}}=-\frac{1}{4}\ket{\mathrm{sAGP}}.
\end{equation}
This shows that the bimodal extreme sAGP is an eigenstate of every bond $H_{pq}$ in the lattice at $\Delta = -1$.

Now we will show it is the ground state at this $\Delta$. Recall the Hamiltonian of the single bond, given in Eqn (\ref{singlebond}). For any states of the entire lattice, only the spin configurations at site $p$ and $q$ have an influence on the single bond, so we can safely project the state to the subspace that only contains these 2 sites and diagonalize the Hamiltonian of the bond in this subspace. The eigenvalues are $-\frac{1}{4},-\frac{1}{4},-\frac{1}{4},\frac{1}{4}$. The bimodal extreme sAGP energy of $-\frac{1}{4}$ means that it is a ground state for this single bond.  Thus bimodal extreme sAGP is a ground state for all bonds in the lattice at $\Delta = -1$. This means it is also a ground state of the entire Hamiltonian, and
\begin{equation}
  H_\mathrm{XXZ}\ket{\mathrm{sAGP}}=-\frac{\text{Number of bonds}}{4}\ket{\mathrm{sAGP}}.
\end{equation}
Note that this result relies only on the form of the Hamiltonian and on the lattice being bipartite.  In particular, it is true for any number of dimensions, for any boundary conditions, and for any (integer) $S^z$ sector.

\subsection{Trimodal Extreme sAGP for Tripartite Lattices}
The kagome and triangular lattices cannot be colored with only 2 colors due to the triangular shape (Fig.\ref{2D lattice}). These lattices are three-colorable (i.e. tripartite). We will show that trimodal extreme sAGP is an eigenstate of the triangular shapes in the three-colorable lattices.

Say $p,q,r$ are 3 sites that form a triangle. In trimodal extreme sAGP, $\eta_p,\eta_q,\eta_r$ are correspondingly $1,e^{\pm\frac{2i\pi}{3}}$. Let 
\begin{equation}
  H_\Delta=H_{pq}+H_{qr}+H_{rp}.
\end{equation}
The trimodal extreme sAGP, when focusing on these three sites, is
\begin{widetext}
\begin{eqnarray}
  &&\ket{\mathrm{sAGP}}\nonumber\\
  &&=\sum_{\sim}c_{\uparrow\uparrow\uparrow}(\sim)\eta_p\eta_q\eta_r\ket{\sim\uparrow_p\uparrow_q\uparrow_r\sim}\nonumber\\
  &&+\sum_{\sim}c_{\uparrow\uparrow\downarrow}(\sim)(\eta_p\eta_q\ket{\sim\uparrow_p\uparrow_q\downarrow_r\sim}+\eta_p\eta_r\ket{\sim\uparrow_p\downarrow_q\uparrow_r\sim}+\eta_r\eta_q\ket{\sim\downarrow_p\uparrow_q\uparrow_r\sim})\nonumber\\
  &&+\sum_{\sim}c_{\uparrow\downarrow\downarrow}(\sim)(\eta_p\ket{\sim\uparrow_p\downarrow_q\downarrow_r\sim}+\eta_q\ket{\sim\downarrow_p\uparrow_q\downarrow_r\sim}+\eta_r\ket{\sim\downarrow_p\downarrow_q\uparrow_r\sim})\nonumber\\
  &&+\sum_{\sim}c_{\downarrow\downarrow\downarrow}(\sim)\ket{\sim\downarrow_p\downarrow_q\downarrow_r\sim}.\nonumber\\
\end{eqnarray}
\end{widetext}

Following a similar procedure as we have outlined for the two-colorable case, it can be shown that
\begin{equation}
  H_\Delta\ket{\mathrm{sAGP}}=-\frac{3}{8}\ket{\mathrm{sAGP}}.
\end{equation}

Thus the trimodal extreme sAGP is the ground of state of a triangle that contains the 3 different $\eta$ values. In periodic boundary conditions, both kagome and triangular lattices are composed purely of such triangles, and trimodal extreme sAGP is the exact ground state at $\Delta = -0.5$. For open boundary conditions, trimodal extreme sAGP is the exact ground state at $\Delta = -0.5$ only when the lattice breaks none of these triangles.

\bibliographystyle{apsrev4-2}
\bibliography{XXZ}

\end{document}